\documentclass[aps,showpacs,twocolumn]{revtex4}
\usepackage{graphicx}

\begin{document}


\title{Thin accretion disks in stationary axisymmetric wormhole spacetimes}

\author{Tiberiu Harko}
\email{harko@hkucc.hku.hk} \affiliation{Department of Physics and
Center for Theoretical and Computational Physics, The University
of Hong Kong, Pok Fu Lam Road, Hong Kong}

\author{Zolt\'{a}n Kov\'{a}cs}
\email{zkovacs@mpifr-bonn.mpg.de}

\affiliation{Max-Planck-Institut f\"{u}r Radioastronomie, Auf dem
H\"{u}gel 69, 53121 Bonn, Germany}
\affiliation{Department of Experimental Physics, University of
Szeged, D\'{o}m T\'{e}r 9, Szeged 6720, Hungary}

\author{Francisco S. N. Lobo}
\email{flobo@cosmo.fis.fc.ul.pt}
\affiliation{Centro de Astronomia e Astrof\'{\i}sica da
Universidade de Lisboa, Campo Grande, Ed. C8 1749-016 Lisboa,
Portugal}

\date{\today}

\begin{abstract}

In this paper, we study the physical properties and the
equilibrium thermal radiation emission characteristics of matter
forming thin accretion disks in stationary axially symmetric
wormhole spacetimes. The thin disk models are constructed by
taking different values of the wormhole's angular velocity, and
the time averaged energy flux, the disk temperature and the
emission spectra of the accretion disks are obtained. Comparing
the mass accretion in a rotating wormhole geometry with the one of
a Kerr black hole, we verify that the intensity of the flux
emerging from the disk surface is greater for wormholes than for
rotating black holes with the same geometrical mass and accretion
rate. We also present the conversion efficiency of the accreting
mass into radiation, and show that the rotating wormholes provide
a much more efficient engine for the transformation of the
accreting mass into radiation than the Kerr black holes. Therefore
specific signatures appear in the electromagnetic spectrum of thin
disks around rotating wormholes, thus leading to the possibility
of distinguishing wormhole geometries by using astrophysical
observations of the emission spectra from accretion disks.

\end{abstract}


\maketitle

\section{Introduction}

In a recent paper, the physical properties and characteristics of
matter forming thin accretion disks in static and spherically
symmetric wormhole spacetimes were extensively analyzed
\cite{Harko:2008vy}. In particular, the time averaged energy flux,
the disk temperature and the emission spectra of the accretion
disks were obtained for these exotic geometries, and compared with
the Schwarzschild solution. It was shown that more energy is
emitted from the disk in a wormhole geometry than in the case of
the Schwarzschild potential. These effects in the disk radiation
were confirmed in the radial profiles of temperature corresponding
to the flux distributions, and in the emission spectrum of the
accretion disks. Thus, specific signatures appear in the
electromagnetic spectrum, leading to the possibility of
distinguishing these static and spherically symmetric wormhole
geometries by using astrophysical observations of the emission
spectra from accretion disks. The results of \cite{Harko:2008vy}
further extended the analysis of the emissivity properties of
accretion disks around general relativistic compact objects
\cite{Bom,To02,YuNaRe04,Guzman:2005bs,Pun:2008ae,Pun:2008ua}.

In the context of stationary axisymmetric spacetimes, the mass
accretion around rotating black holes was studied in general
relativity for the first time in \cite{NoTh73}. By using an
equatorial approximation to the stationary and axisymmetric
spacetime of rotating black holes, steady-state thin disk models
were constructed, extending the theory of nonrelativistic
accretion \cite{ShSu73}. In these models hydrodynamical
equilibrium is maintained by efficient cooling mechanisms via
radiation transport, and the accreting matter has a Keplerian
rotation. The radiation emitted by the disk surface was also
studied under the assumption that black body radiation would
emerge from the disk in thermodynamical equilibrium. The radiation
properties of the thin accretion disks were further analyzed  in
 \cite{PaTh74,Th74}, where the effects of the photon
capture by the black hole on the spin evolution were presented as
well. In these works the efficiency with which black holes convert
rest mass into outgoing radiation in the accretion process was
also computed.

It is of interest to consider the properties of thin accretion
disks around wormhole geometries, as these exotic spacetimes
violate the null energy condition (NEC) \cite{Harko:2008vy}.
Indeed, a wide variety of wormhole solutions have been considered
in the literature (we refer the reader to
\cite{WHsolutions,Lobo:2007zb} and to \cite{Lobo:2007zb} for a
recent review). It will also prove interesting to further extend
the analysis carried out in \cite{Harko:2008vy} to stationary
axisymmetric wormhole geometries \cite{teo}, which are physically
more realistic objects than their static and spherically symmetric
counterparts. It is also important to compare the properties of
the thin accretion disks around rotating wormholes with the
properties of thin disks around rotating black holes. From this
comparison it follows that the intensity of the flux emerging from
the disk surface is greater for wormholes than for the Kerr black
holes with the same geometrical mass and accretion rate. This
gives an effective observational method to discriminate between
wormholes and black hole type compact general relativistic
objects.

The present paper is organized as follows. In Sec. \ref{sec:II},
we review the formalism and the physical properties of the thin
disk accretion onto compact objects, for stationary axisymmetric
spacetimes. In Sec. \ref{sec:III}, we analyze the basic properties
of matter forming a thin accretion disk in rotating wormhole
spacetimes. We discuss and conclude our results in Sec.
\ref{sec:concl}. In the Appendix, we present for self-completeness
and self-consistency the effective potential for the Kerr black
hole. Throughout this work, we use a system of units so that
$c=G=\hbar =k_{B}=1$, where $k_{B}$ is Boltzmann's constant.

\section{Thermal equilibrium radiation properties of thin accretion
disks in stationary axisymmetric spacetimes}\label{sec:II}

\subsection{Stationary and axially symmetric spacetimes}

The physical properties and the electromagnetic radiation
characteristics of particles moving in circular orbits around
general relativistic bodies are determined by the geometry of the
spacetime around the compact object.  For a stationary and axially
symmetric geometry the metric is given in a general form by
\begin{equation}\label{rotmetr1}
ds^2=g_{tt}\,dt^2+2g_{t\phi}\,dt d\phi+g_{rr}\,dr^2
+g_{\theta\theta}\,d\theta^2+g_{\phi\phi}\,d\phi^2\,.
\end{equation}
In the equatorial approximation, which is the case of interest for
our analysis, the metric functions $g_{tt}$, $g_{t\phi}$,
$g_{rr}$, $g_{\theta\theta}$ and $g_{\phi\phi}$ only depend on the
radial coordinate $r$, i.e., $|\theta-\pi /2|\ll 1$.

To compute the relevant physical quantities of thin accretion
disks, we determine first the radial dependence of the angular
velocity $\Omega $, of the specific energy $\widetilde{E}$, and of
the specific angular momentum $\widetilde{L}$ of particles moving
in circular orbits in a stationary and axially symmetric geometry
through the geodesic equations. The latter take the following form
\begin{eqnarray}
\frac{dt}{d\tau}&=&\frac{\widetilde{E}
g_{\phi\phi}+\widetilde{L}g_{t\phi}}{g_{t\phi}^2-g_{tt}g_{\phi\phi}},
   \label{geodeqs1}   \\
\frac{d\phi}{d\tau}&=&-\frac{\widetilde{E}
g_{t\phi}+\widetilde{L}g_{tt}}{g_{t\phi}^2-g_{tt}g_{\phi\phi}},
    \label{geodeqs2}  \\
g_{rr}\left(\frac{dr}{d\tau}\right)^2&=&-1+\frac{\widetilde{E}^2
g_{\phi\phi}+2\widetilde{E}\widetilde{L}g_{t\phi}
+\widetilde{L}^2g_{tt}}{g_{t\phi}^2-g_{tt}g_{\phi\phi}}.
    \label{geodeqs3}
\end{eqnarray}
From Eq.~(\ref{geodeqs3}) one can introduce an effective potential
term as
\begin{equation}\label{roteffpot}
V_{eff}(r)=-1+\frac{\widetilde{E}^2
g_{\phi\phi}+2\widetilde{E}\widetilde{L}g_{t\phi}
+\widetilde{L}^2g_{tt}}{g_{t\phi}^2-g_{tt}g_{\phi\phi}}.
\end{equation}

For stable circular orbits in the equatorial plane the following
conditions must hold: $V_{eff}(r)=0$ and $V_{eff,\;r}(r)=0$, where
the comma in the subscript denotes a derivative with respect to
the radial coordinate $r$. These conditions provide the specific
energy, the specific angular momentum and the angular velocity of
particles moving in circular orbits for the case of spinning
general relativistic compact spheres, given by
\begin{eqnarray}
\widetilde{E}&=&-\frac{g_{tt}+g_{t\phi}\Omega}{\sqrt{-g_{tt}
-2g_{t\phi}\Omega-g_{\phi\phi}\Omega^2}},
    \label{rotE}  \\
\widetilde{L}&=&\frac{g_{t\phi}+g_{\phi\phi}\Omega}{\sqrt{-g_{tt}
-2g_{t\phi}\Omega-g_{\phi\phi}\Omega^2}},
     \label{rotL}  \\
\Omega&=&\frac{d\phi}{dt}=\frac{-g_{t\phi,r}+\sqrt{(g_{t\phi,r})^2
-g_{tt,r}g_{\phi\phi,r}}}{g_{\phi\phi,r}}.
     \label{rotOmega}
\end{eqnarray}
The marginally stable orbit around the central object can be
determined from the condition $V_{eff,\;rr}(r)=0$. To this effect,
we formally represent the effective potential as
\[
V_{eff}(r)\equiv-1+\frac{f}{g},
\]
where
\begin{eqnarray*}
f & \equiv & \widetilde{E}^{2}g_{\phi\phi}
+2\widetilde{E}\widetilde{L}g_{t\phi}+\widetilde{L}^{2}g_{\phi\phi},\\
g & \equiv & g_{t\phi}^{2}-g_{tt}g_{\phi\phi},
\end{eqnarray*}
and where the condition $g\neq 0$ is imposed. From $V_{eff}(r)=0$,
we obtain first $f=g$. The condition $V_{eff,\;r}(r)=0$ provides
$f_{,r}g-fg_{,r}=0$. Thus, from these conditions one readily
derives $V_{eff,\;rr}(r)=0$, which provides the following
important relationship
\begin{eqnarray}
 0& = & (g_{t\phi}^2-g_{tt}g_{\phi\phi})V_{eff,rr}\nonumber\\
 & = &  \widetilde{E}^{2}g_{\phi\phi,rr}+2\widetilde{E}\widetilde{L}g_{t\phi,rr} +\widetilde{L}^{2}g_{tt,rr}  \nonumber \\
& & -(g_{t\phi}^{2} -g_{tt}g_{\phi\phi})_{,rr}\;,\label{mso-r}
\end{eqnarray}
where $g_{t\phi}^2-g_{tt}g_{\phi\phi}$ (appearing as a cofactor in
the metric determinant) never vanishes. By inserting
Eqs.~(\ref{rotE})-(\ref{rotOmega}) into Eq.~(\ref{mso-r}) and
solving this equation for $r$, we obtain the radii of the
marginally stable orbits, once the metric coefficients $g_{tt}$,
$g_{t\phi}$ and $g_{\phi\phi}$ are explicitly given.

\subsection{Physical properties of thin accretion disks}

For a thin accretion disk the vertical size (defined in
cylindrical coordinates along the $z$-axis) is negligible, as
compared to its horizontal extension (defined along the radial
direction $r$), i.e., the disk height $H$, equal to the maximum
half thickness of the disk, is always much smaller than the
characteristic radius $R$ of the disk, $H \ll R$.  The thin disk
is assumed to be in hydrodynamical equilibrium, and the pressure
gradient, as well as the vertical entropy gradient, are negligible
in the disk. The efficient cooling via the radiation over the disk
surface prevents the disk from cumulating the heat generated by
stresses and dynamical friction. In turn, this equilibrium causes
the disk to stabilize its thin vertical size. The thin disk has an
inner edge at the marginally stable orbit of the compact object
potential, and the accreting matter has a Keplerian motion in
higher orbits.

In steady-state accretion disk models, the mass accretion rate
$\dot{M}_{0}$ is assumed to be a constant that does not change
with time. The physical quantities describing the orbiting matter
are averaged over a characteristic time scale, e.g. $\Delta t$,
for a total period of the orbits, over the azimuthal angle $\Delta
\phi =2\pi $,  and over the height $H$
\cite{ShSu73,NoTh73,PaTh74}.

The particles moving in Keplerian orbits around the compact object
with a rotational velocity $\Omega =d\phi /dt$ have a specific
energy $\widetilde{E} $ and a specific angular momentum
$\widetilde{L\text{,}}$ which in the steady-state thin disk model
depend only on the radii of the orbits. These particles, orbiting
with the four-velocity $u^{\mu }$, form a disk of an averaged
surface density $\Sigma $, the vertically integrated average of
the rest mass density $\rho _{0}$ of the plasma. The accreting
matter in the disk is modeled by an anisotropic fluid source,
where the density $\rho _{0}$, the energy flow vector $q^{\mu }$
and the stress tensor $t^{\mu \nu }$ are measured in the averaged
rest-frame (the specific heat was neglected). Then, the disk
structure can be characterized by the surface density of the disk
\cite{NoTh73,PaTh74}
\begin{equation}
\Sigma(r) = \int^H_{-H}\langle\rho_0\rangle dz,
\end{equation}
with averaged rest mass density $\langle\rho_0\rangle$ over
$\Delta t$ and $ 2\pi$ and the torque
\begin{equation}
W_{\phi}{}^{r} =\int^H_{-H}\langle t_{\phi}{}^{r}\rangle dz,
\end{equation}
with the averaged component $\langle t^r_{\phi} \rangle$ over
$\Delta t$ and $2\pi$. The time and orbital average of the energy
flux vector gives the radiation flux ${\mathcal F}(r)$ over the
disk surface as ${\mathcal F}(r)=\langle q^z \rangle$.

The stress-energy tensor is decomposed according to
\begin{equation}
T^{\mu \nu }=\rho_{0}u^{\mu }u^{\nu }+2u^{(\mu }q^{\nu )}+t^{\mu
\nu },
\end{equation}
where $u_{\mu }q^{\mu }=0$, $u_{\mu }t^{\mu \nu }=0$. The
four-vectors of the energy and angular momentum flux are defined
by $-E^{\mu }\equiv T_{\nu }^{\mu }{}(\partial /\partial t)^{\nu
}$ and $J^{\mu }\equiv T_{\nu }^{\mu }{}(\partial /\partial \phi
)^{\nu }$, respectively. The structure equations of the thin disk
can be derived by integrating the conservation laws of the rest
mass, of the energy, and of the angular momentum of the plasma,
respectively \cite{NoTh73,PaTh74}. From the equation of
the rest mass conservation, $\nabla _{\mu }(\rho _{0}u^{\mu })=0$,
 it follows that the time averaged rate of the accretion of the
rest mass is independent of the disk radius,
\begin{equation}
\dot{M_{0}}\equiv -2\pi r\Sigma u^{r}={\rm constant}.
\label{conslawofM}
\end{equation}

The conservation law $\nabla _{\mu }E^{\mu }=0$ of the energy has
the integral form
\begin{equation}
\lbrack \dot{M}_{0}\widetilde{E}-2\pi r\Omega W_{\phi }{}^{r}]_{,r}
=4\pi r{\mathcal F}%
\widetilde{E},  \label{conslawofE}
\end{equation}
which states that the energy transported by the rest mass flow,
$\dot{M}_{0} \widetilde{E}$, and the energy transported by the
dynamical stresses in the disk, $2\pi r\Omega W_{\phi }{}^{r}$, is
in balance with the energy radiated away from the surface of the
disk, $4\pi r{\mathcal F}\widetilde{E}$. The law of the angular
momentum conservation, $\nabla _{\mu }J^{\mu }=0$, also states the
balance of these three forms of the angular momentum transport,
\begin{equation}
\lbrack \dot{M}_{0}\widetilde{L}-2\pi rW_{\phi }{}^{r}]_{,r}=4\pi
r{\mathcal F} \widetilde{L}.  \label{conslawofL}
\end{equation}

By eliminating $W_{\phi }{}^{r}$ from Eqs.~(\ref{conslawofE}) and
(\ref {conslawofL}), and applying the universal energy-angular
momentum relation $ dE=\Omega dJ$ for circular geodesic orbits in
the form $\widetilde{E}_{,r}=\Omega \widetilde{L}_{,r}$, the flux
${\mathcal F}$ of the radiant energy over the disk can be
expressed in terms of the specific energy, angular momentum and of
the angular velocity of the compact sphere \cite{NoTh73,PaTh74},
\begin{equation}
{\mathcal F}(r)=-\frac{\dot{M}_{0}}{4\pi \sqrt{-g}}\frac{\Omega
_{,r}}{(\widetilde{E}-\Omega
\widetilde{L})^{2}}\int_{r_{ms}}^{r}(\widetilde{E}-\Omega
\widetilde{L}) \widetilde{L}_{,r}dr.  \label{F}
\end{equation}

Another important characteristic of the mass accretion process is
the efficiency with which the central object converts rest mass
into outgoing radiation. This quantity is defined as the ratio of
the rate of the radiation of the energy of photons escaping from
the disk surface to infinity, and the rate at which mass-energy is
transported to the central compact general relativistic object,
both measured at infinity \cite{NoTh73,PaTh74}. If all the emitted
photons can escape to infinity, the efficiency is given in terms
of the specific energy measured at the marginally stable orbit
$r_{ms}$,
\begin{equation}
\epsilon =1-\widetilde{E}_{ms}.\label{epsilon}
\end{equation}

For Schwarzschild black holes the efficiency $\epsilon $ is about
$6\%$, whether the photon capture by the black hole is considered,
or not. Ignoring the capture of radiation by the hole, $\epsilon $
is found to be $42\%$ for rapidly rotating black holes, whereas
the efficiency is $40\%$ with photon capture in the Kerr potential
\cite{Th74}.

The accreting matter in the steady-state thin disk model is
supposed to be in thermodynamical equilibrium. Therefore the
radiation emitted by the disk surface can be considered as a
perfect black body radiation, where the energy flux is given by
${\mathcal F}(r)=\sigma T^{4}(r)$ ($\sigma $ is the
Stefan-Boltzmann constant), and the observed luminosity $L\left(
\nu \right) $ has a redshifted black body spectrum \citep{To02}:
\begin{equation}
L\left( \nu \right) =4\pi d^{2}I\left( \nu \right)
=\frac{8}{\pi }\cos \gamma \int_{r_{i}}^{r_{f}}
\int_0^{2\pi}\frac{\nu^{3}_e r d\phi dr }{\exp \left( h\nu_e/T\right) -1}.
\end{equation}

Here $d$ is the distance to the source, $I(\nu )$ is the Planck
distribution function, $\gamma $ is the disk inclination angle,
and $r_{i}$ and $r_{f}$ indicate the position of the inner and
outer edge of the disk, respectively. We take $r_{i}=r_{ms}$ and
$r_{f}\rightarrow \infty $, since we expect the flux over the disk
surface vanishes at $r\rightarrow \infty $ for any kind of general
relativistic compact object geometry. The emitted frequency is
given by $\nu_e=\nu(1+z)$, and the redshift factor can be written
as
\begin{equation}
1+z=\frac{1+\Omega r \sin \phi \sin \gamma }{\sqrt{ -g_{tt}
- 2 \Omega g_{t\phi} - \Omega^2 g_{\phi\phi}}},
\end{equation}
where we have neglected the light bending \citep{Lu79,BMT01}.

The flux and the emission spectrum of the accretion disks around
compact objects satisfy some simple scaling relations, with
respect to the simple scaling transformation of the radial
coordinate, given by $r\rightarrow \widetilde{r}=r/ M$, where $M$
is the mass of the compact sphere. Generally, the metric tensor
coefficients are invariant with respect of this transformation,
while the specific energy, the angular momentum and the angular
velocity transform as $ \widetilde{E}\rightarrow \widetilde{E}$,
$\widetilde{L}\rightarrow M\widetilde{L
}$ and $%
\Omega \rightarrow \widetilde{\Omega}/M$, respectively. The flux
scales as $F(r) \rightarrow F( \widetilde{r})/M^{4}$, giving the
simple transformation law of the temperature as $ T(r)\rightarrow
T\left( \widetilde{r}\right) /M$. By also rescaling the frequency
of the emitted radiation as  $\nu \rightarrow \widetilde{\nu}=\nu
/M$, the luminosity of the disk is given by $L\left( \nu \right)
\rightarrow L\left( \widetilde{\nu}\right) /M$. On the other hand,
the flux is proportional to the accretion rate $\dot{M}_{0}$, and
therefore an increase in the accretion rate leads to a linear
increase in the radiation emission flux from the disk.

\section{Accretion disk properties in rotating
wormhole geometries}\label{sec:III}

\subsection{Metric and field equations}

The canonical metric for a stationary, axisymmetric traversable
wormhole can be written as \cite{teo}
\begin{equation}\label{3rwh}
ds^2=-N^2dt^2+e^{\mu}\,dr^2+r^2K^2[d\theta^2
+\sin^2\theta(d\phi-\omega\,dt)^2]\;,  \label{rotWH}
\end{equation}
where $N$, $\mu$, $K$ and $\omega$ are functions of $r$ and
$\theta$. To ensure that the metric is nonsingular on the rotation
axis ($\theta=0,\pi$), regularity conditions on $N$, $\mu$ and $K$
have to be imposed \cite{teo}, which essentially means that their
respective $\theta$ derivatives have to vanish on the rotation
axis.

For simplicity, we shall consider the following definitions
\cite{teo}
\begin{equation}
N(r,\theta)=e^{\Phi(r,\theta)}\,,\qquad
e^{-\mu(r,\theta)}=1-\frac{b(r,\theta)}{r}\,,
\end{equation}
which are well suited to describe a traversable wormhole.
$\Phi(r,\theta)$ is the redshift function, which needs to be
finite to ensure that there are no event horizons or curvature
singularities. $b(r,\theta)$ is the shape function which satisfies
$b\leq r$ and the flaringout condition. $K(r,\theta)$ determines
the proper radial distance, while $\omega $ governs the angular
velocity of the wormhole.

The scalar curvature of the space-time (\ref{3rwh}) is extremely
messy, but at the throat $r=r_0$ simplifies to
\begin{eqnarray}\label{rotWHRicciscalar}
R&=&-\frac{1}{r^2K^2}\left(\mu_{\theta\theta}
+\frac{1}{2}\mu_\theta^2\right)
-\frac{\mu_\theta}{Nr^2K^2}\,\frac{(N
\sin\theta)_\theta}{\sin\theta}
    \nonumber    \\
&& -\frac{2}{Nr^2K^2}\,\frac{(N_{\theta}
\sin\theta)_\theta}{\sin\theta}
    -\frac{2}{r^2K^3}\,\frac{(K_\theta
\sin\theta)_\theta}{\sin\theta}
           \nonumber    \\
&&+e^{-\mu}\,\mu_r\,\left[\ln(Nr^2K^2)\right]_r
+\frac{\sin^2\theta\,\omega_\theta^2}{2N^2}
   \nonumber    \\
&&+\frac{2}{r^2K^4}\,(K^2+K_\theta^2)  \,,
\end{eqnarray}
where the subscripts denote partial derivatives with respect to
$r$ and $\theta$. Note that the only troublesome terms are the
ones involving $\mu_\theta$ and $\mu_{\theta\theta}$, i.e.,
\begin{equation}
\mu_\theta=\frac{b_\theta}{(r-b)}\,,   \qquad \mu_{\theta\theta}
+\frac{1}{2}\mu_\theta^2=\frac{b_{\theta\theta}}{r-b}
+\frac{3}{2}{b_\theta{}^2\over(r-b)^2}\,.
\end{equation}
Thus, one needs to impose that $b_\theta=0$ and
$b_{\theta\theta}=0$ at the throat to avoid curvature
singularities. This condition shows that the throat is located at
a constant value of $r$.

The stress-energy tensor components are extremely complicated, but
assume a more simplified form in an orthonormal reference frame
and evaluated at the throat. They have the following nonzero
components
\begin{widetext}
\vspace{-1cm}
\begin{eqnarray}
8\pi T_{\hat{t}\hat{t}}\big|_{r=r_0}&=&-\frac{(K_\theta
\sin\theta)_\theta}{r^2K^3\sin\theta}
-\frac{\omega_\theta^2\,\sin^2\theta}{4N^2}
+e^{-\mu}\,\mu_r\,\frac{(rK)_r}{rK}
       +\frac{K^2+K_\theta^2}{r^2K^4}
    \,,   \label{rotGtt}
\\
8\pi T_{\hat{r}\hat{r}}\big|_{r=r_0}&=&\frac{(K_\theta
\sin\theta)_\theta}{r^2K^3\sin\theta}
-\frac{\omega_\theta^2\,\sin^2\theta}{4N^2}
+\frac{(N_\theta \sin\theta)_\theta}{Nr^2K^2\sin\theta}
-\frac{K^2+K_\theta^2}{r^2K^4}
   \,,
\\
8\pi T_{\hat{\theta}\hat{\theta}}\big|_{r=r_0}&=& \frac{N_\theta(K
\sin\theta)_\theta}{Nr^2K^3\sin\theta}
+\frac{\omega_\theta^2\,\sin^2\theta}{4N^2}
-\frac{\mu_r\,e^{-\mu}(NrK)_r}{2NrK}
   \,,
\\
8\pi T_{\hat{\phi}\hat{\phi}}\big|_{r=r_0}&=&
-\frac{\mu_r\,e^{-\mu}\,(NKr)_r}{2NKr}-\frac{3\sin^2\theta\,
\omega_\theta^2}{4N^2}
+\frac{N_{\theta\theta}}{Nr^2K^2}-\frac{N_{\theta}K_{\theta}}{Nr^2K^3}
\,,    \label{rotGphiphi}
\\
8\pi
T_{\hat{t}\hat{\phi}}\big|_{r=r_0}&=&\frac{1}{4N^2K^2r}\;\Big(6NK\,
\omega_{\theta}\,\cos\theta
+2NK\,\sin \theta\,\omega_{\theta \theta}
       \nonumber     \\
&& -\mu_{r}e^{-\mu}r^2NK^3\,\sin\theta\; \omega_{r}
+4N\,\omega_\theta\,\sin\theta\,K_\theta
-2K\,\sin\theta\,N_{\theta}\,\omega_{\theta}  \Big)   \,.
           \label{rotGtphi}
\end{eqnarray}

\end{widetext}
The components $T_{\hat{t}\hat{t}}$ and $T_{\hat{i}\hat{j}}$ have
the usual physical interpretations, and in particular,
$T_{\hat{t}\hat{\phi}}$ characterizes the rotation of the matter
distribution.

Taking into account the stress-energy tensor components above, the
NEC at the throat is given by
\begin{eqnarray}\label{NEC}
8\pi\,T_{\hat{\mu} \hat{\nu}}k^{\hat{\mu}}
k^{\hat{\nu}}\big|_{r=r_0}&=&{\rm e}^{-\mu}\mu_r{(rK)_r\over rK}
-{\omega_\theta{}^2\sin^2\theta\over2N^2}
\nonumber  \\
&& +{(N_\theta\sin\theta)_\theta\over(rK)^2N\sin\theta}\,.
\end{eqnarray}
Rather than reproduce the analysis here, we refer the reader to
Ref.~\cite{teo} for details, where it was shown that the NEC is
violated in certain regions, and is satisfied in others. Thus, it
is possible for an infalling observer to move around the throat,
and avoid the exotic matter supporting the wormhole. However, it
is important to emphasize that one cannot avoid the use of exotic
matter altogether.

\subsection{Electromagnetic signatures of thin accretion disks in
rotating wormhole geometries}

The radial geodesic equation (\ref{geodeqs3}) for the metric
(\ref{rotWH}) is given by
\begin{eqnarray}
\left(1-\frac{b}{r}\right)^{-1}\left(\frac{dr}{d\tau}\right)^{2}
=V_{eff} \;.
\end{eqnarray}
Using the relationship
$g_{t\phi}^{2}-g_{tt}g_{\phi\phi}=r^{2}K^{2}e^{2\Phi}$, the
effective potential takes the following form
\begin{eqnarray}
V_{eff}(r)&=&-1+\Big[\widetilde{E}^{2}r^{2}K^{2}
-2\widetilde{E}\widetilde{L}r^{2}K^{2}\omega
  \nonumber    \\
&&-\widetilde{L}^{2}\left(e^{2\Phi}-r^{2}K^{2}\omega^{2}\right)\Big]\big/
\left(r^{2}K^{2}e^{2\Phi}\right)\,.
\end{eqnarray}

In analogy to the effective potential for a Kerr black hole,
provided by Eq.~(\ref{KerrPot}), the above relationships may be
rewritten in the following manner
\begin{equation}
r^4\left(\frac{dr}{d\tau}\right)^{2}=V(r)
  \label{WHPot}
\end{equation}
with $V(r)$ given by
\begin{equation}
V(r)=r^{4}\left(1-\frac{b}{r}\right) V_{eff}(r)\;.\label{WHVeff}
\end{equation}

In this work, we consider the specific case of
\begin{equation}
\omega={2J\over r^3} \,,
\end{equation}
where $J=M^2a_*$ is the total angular momentum of the wormhole. As
we are only interested in the equatorial approximation, i.e.,
$|\theta-\pi/2|\ll 1$, we also consider $K=1$ throughout our
analysis.

We consider the case of $\Phi=- r_0/r$, and the following
respective shape functions
\begin{equation}
b=r_0\,, \qquad  b(r)=\frac{r_0^2}{r}\,, \qquad
b(r)=\sqrt{rr_0}\,,
   \label{form1}
\end{equation}
and
\begin{equation}
b(r)=r_0+\gamma r_0\left(1-\frac{r_0}{r}\right)\,,
   \label{form2}
\end{equation}
with $0<\gamma<1$.

For simplicity, we also assume the following values for the mass
and the spin parameter
\begin{eqnarray}
M&=&0.06776 M_{\odot } (=1000 \;{\rm cm})\\
a_*&=&0.2,\; 0.4,\; 0.6,\; 0.8,\;1.0\,
\end{eqnarray}
respectively.

\begin{figure*}[t]
\centering
 \includegraphics[width=2.8in]{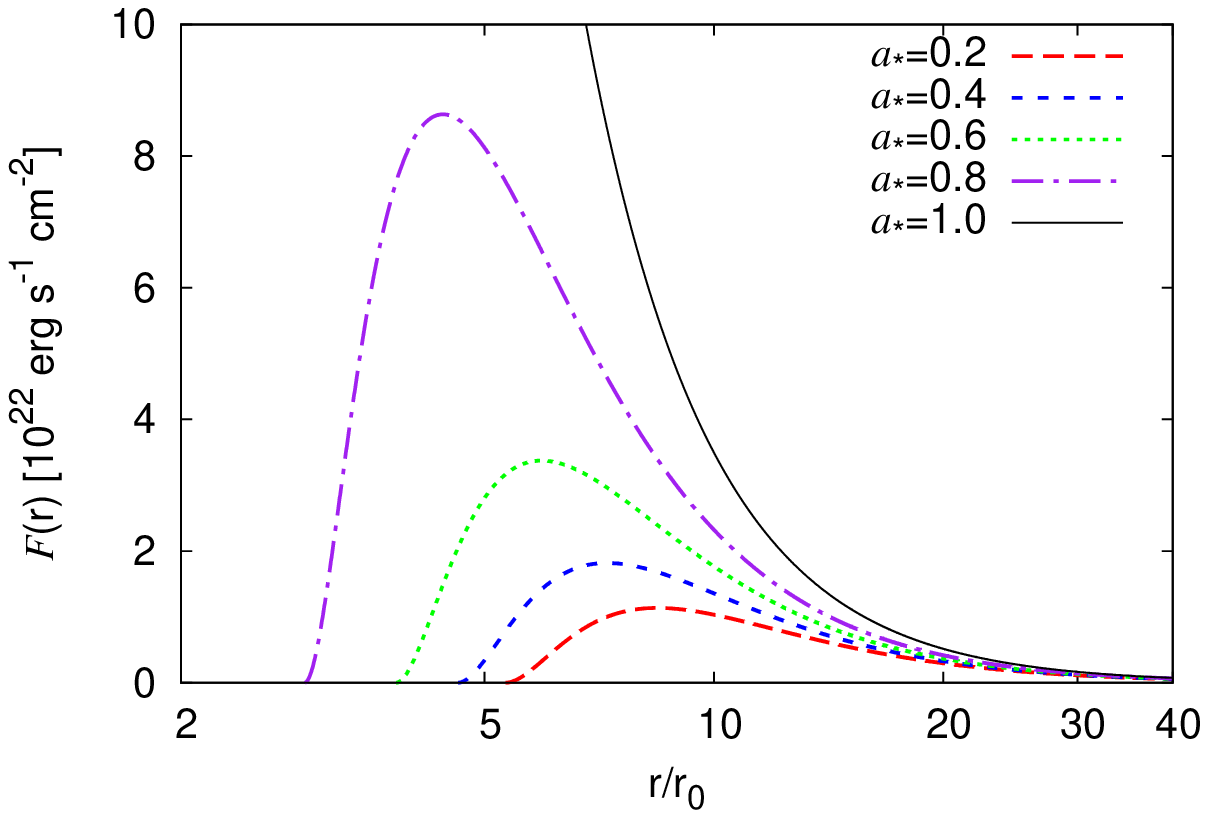}
\hspace{0.2in}
\includegraphics[width=2.8in]{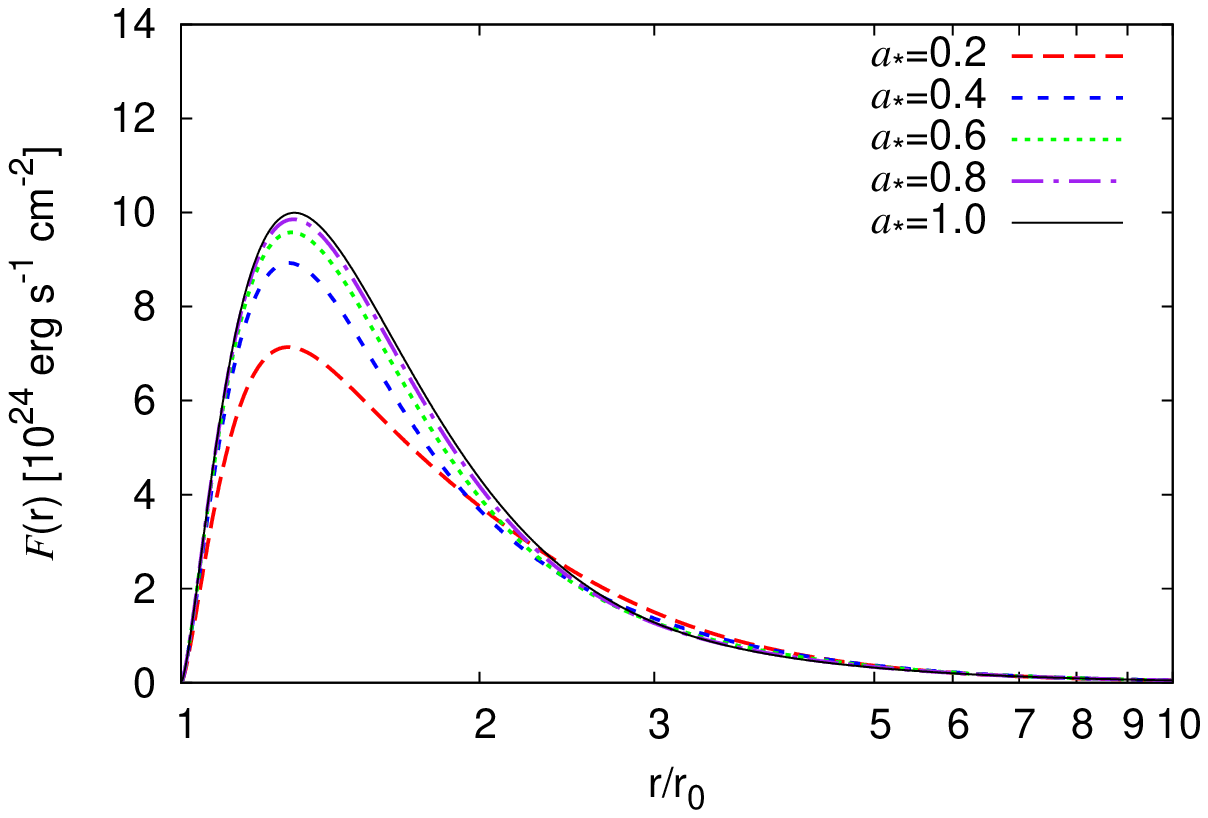}
\hspace{0.2in}
\includegraphics[width=2.8in]{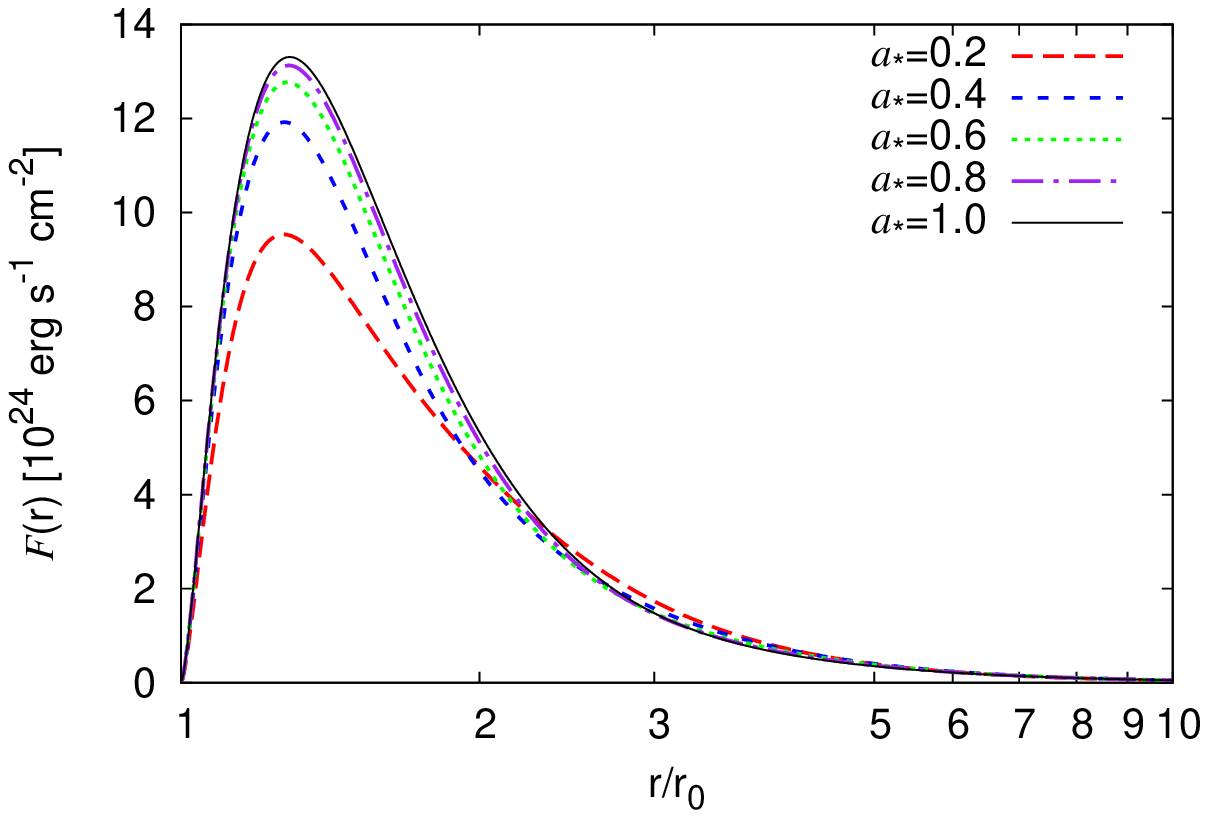}
\hspace{0.2in}
\includegraphics[width=2.8in]{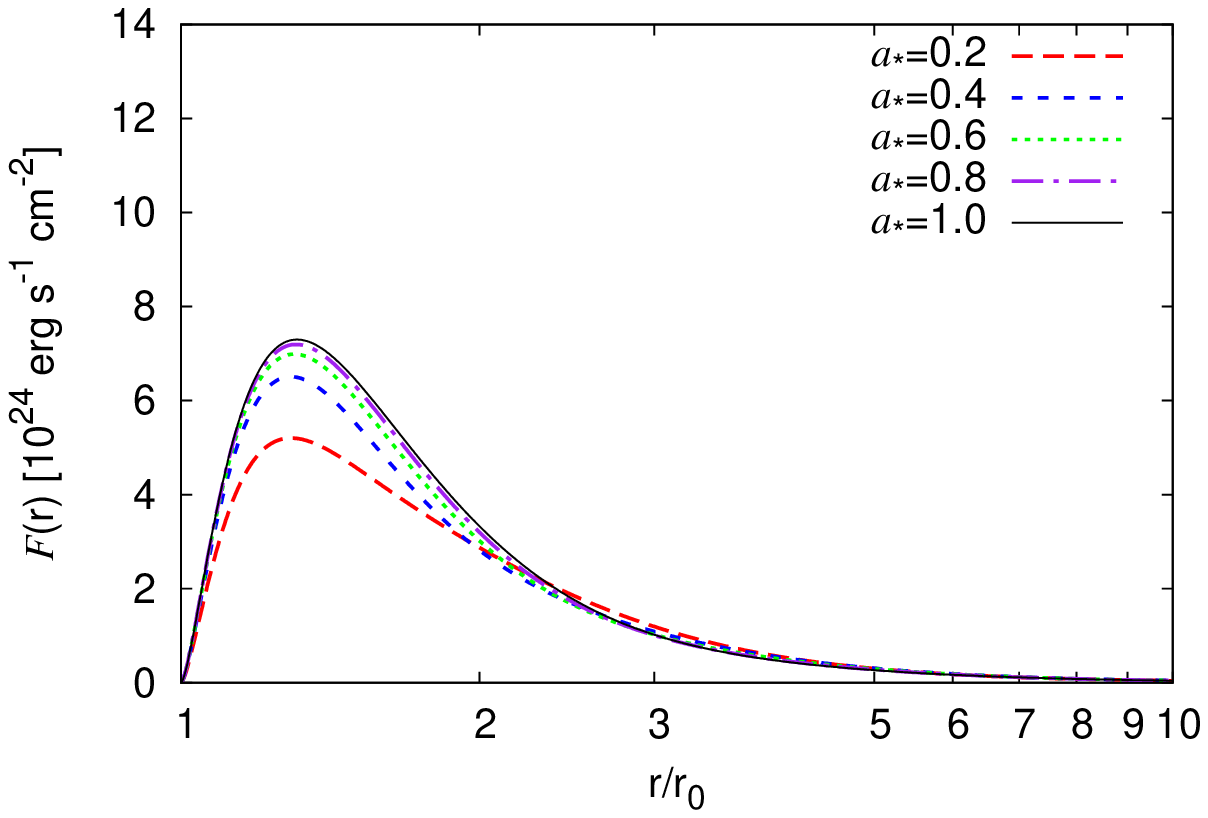}
\caption{The energy flux radiated by an accretion disk in a Kerr
black hole geometry (upper left hand plot), and in the stationary
axially symmetric wormhole spacetimes for $\Phi=- r_0/r$ and
$b=r_0$ (upper right hand plot), $b=r^2_0/r$ (lower left hand
plot) and $b = (r_0 r)^{1/2}$ (lower right hand plot),
respectively. In all plots $r_0 =1000$ cm, and the values of the
spin parameter are $a_* = 0.2, 0.4, 0.6, 0.8$ and 1,
respectively.}
 \label{Fig:rotwh-flux}
\end{figure*}

\begin{figure*}
\centering
\includegraphics[width=2.8in]{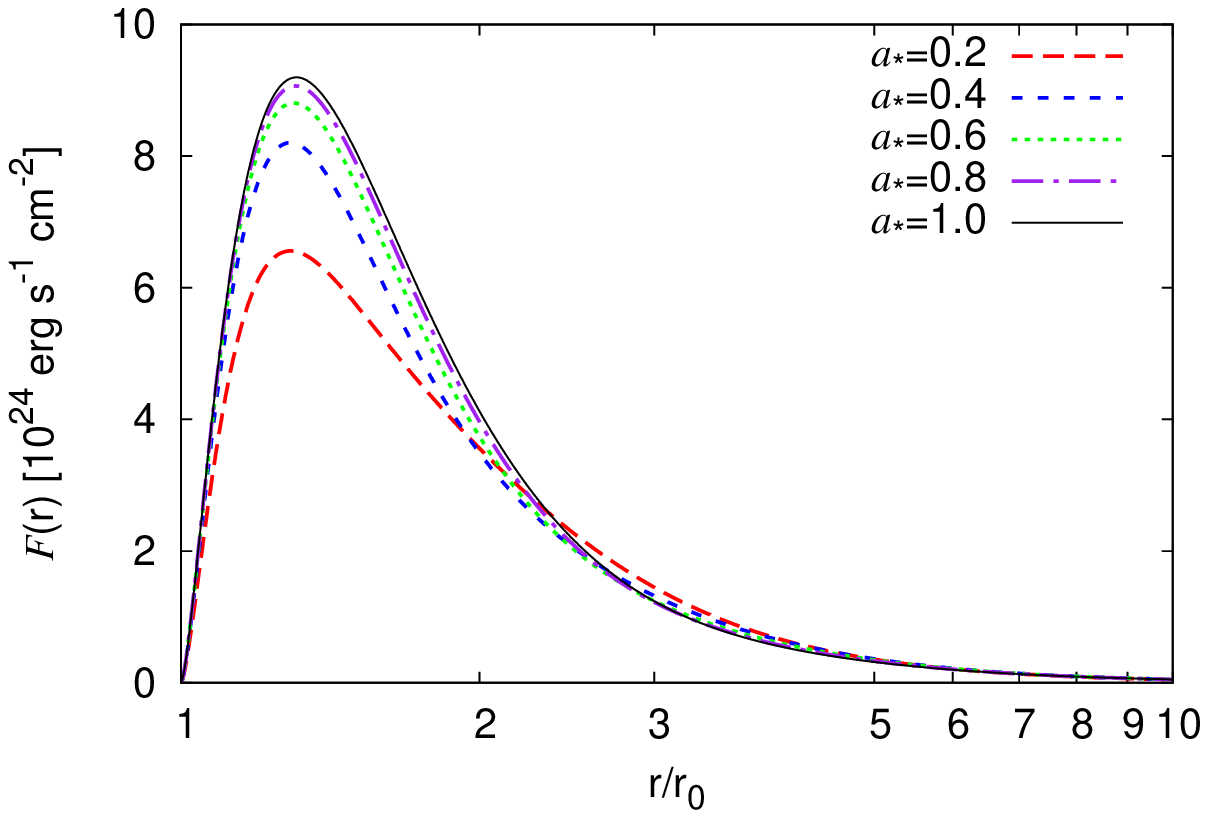}
\hspace{0.2in}
\includegraphics[width=2.8in]{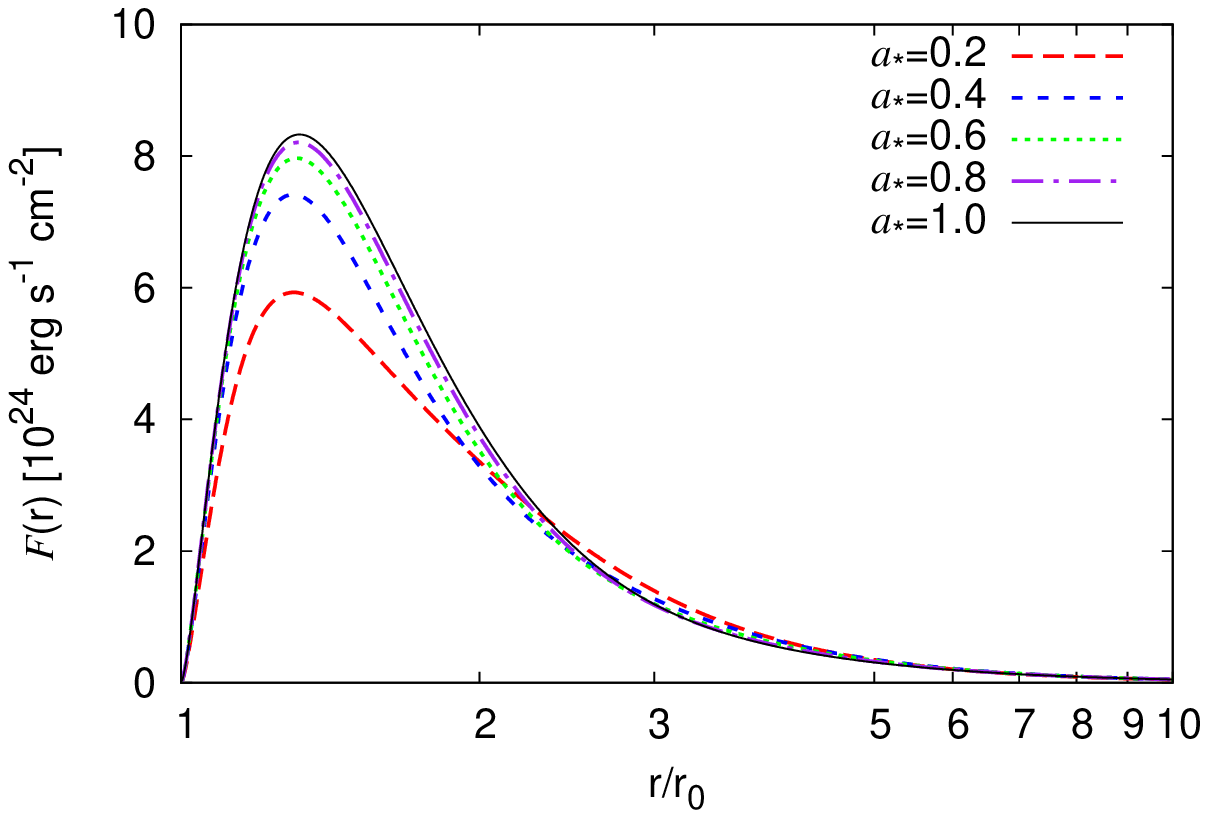}
\hspace{0.2in}
\includegraphics[width=2.8in]{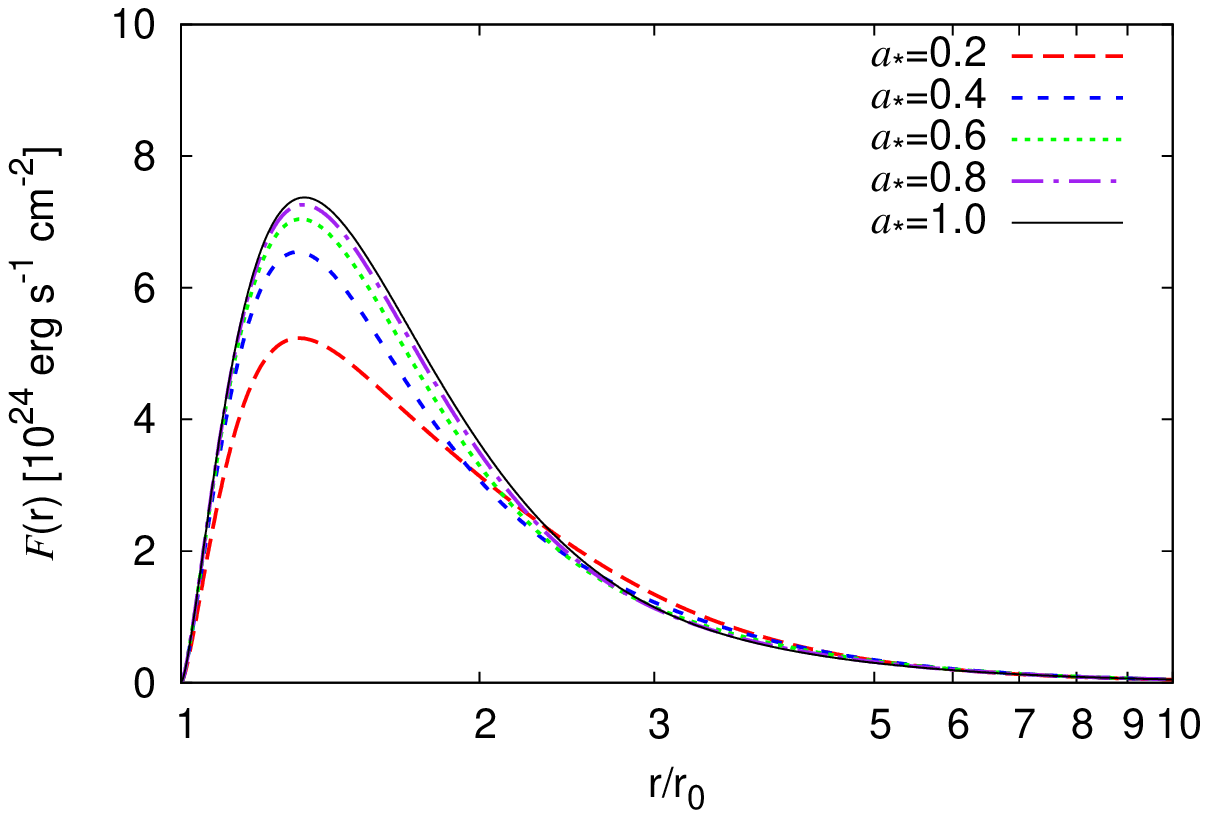}
\hspace{0.2in}
\includegraphics[width=2.8in]{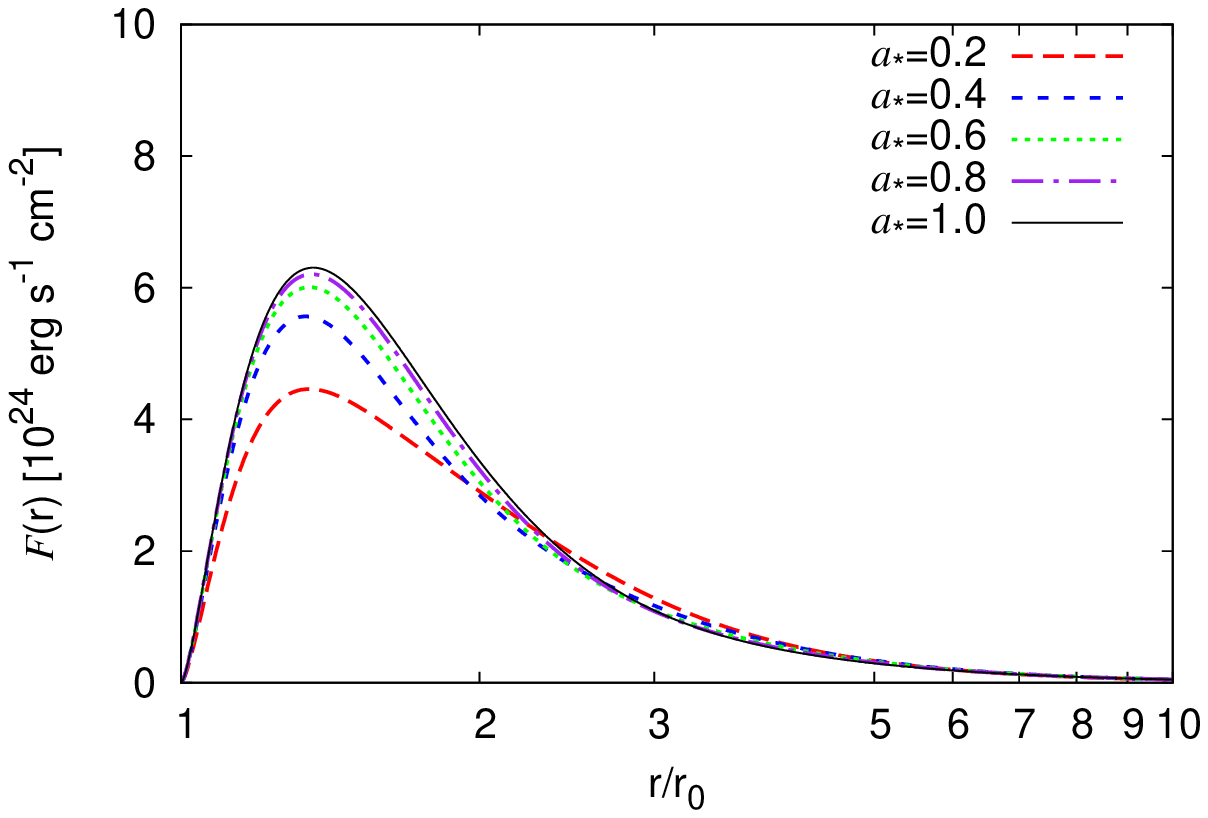}
\caption{The energy flux from accretion disks in stationary
axially symmetric wormhole spacetimes for $\Phi=- r_0/r$, and
$b=r_0+\gamma r_0(1-r_0/r)$, where $\gamma=0.2$ (upper left hand
plot), $0.4$ (upper right hand plot), $0.6$ (lower left hand plot)
and $0.8$ (lower right hand plot), respectively. In all plots $r_0
=1000$ cm, and the values of the spin parameter are $a_* = 0.2,
0.4, 0.6, 0.8$ and 1, respectively.}
 \label{Fig:rotwh-flux2}
\end{figure*}

\begin{figure*}
\centering
\includegraphics[width=2.8in]{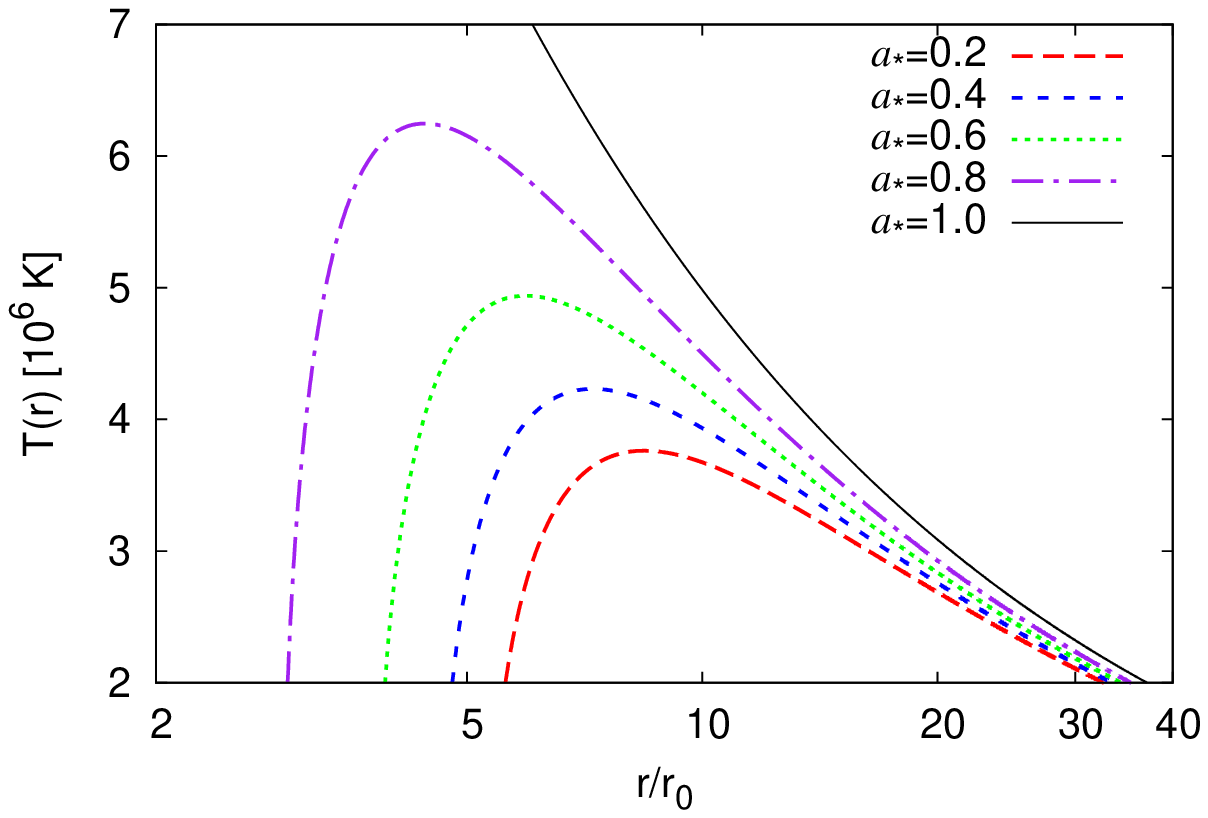}
\hspace{0.2in}
\includegraphics[width=2.8in]{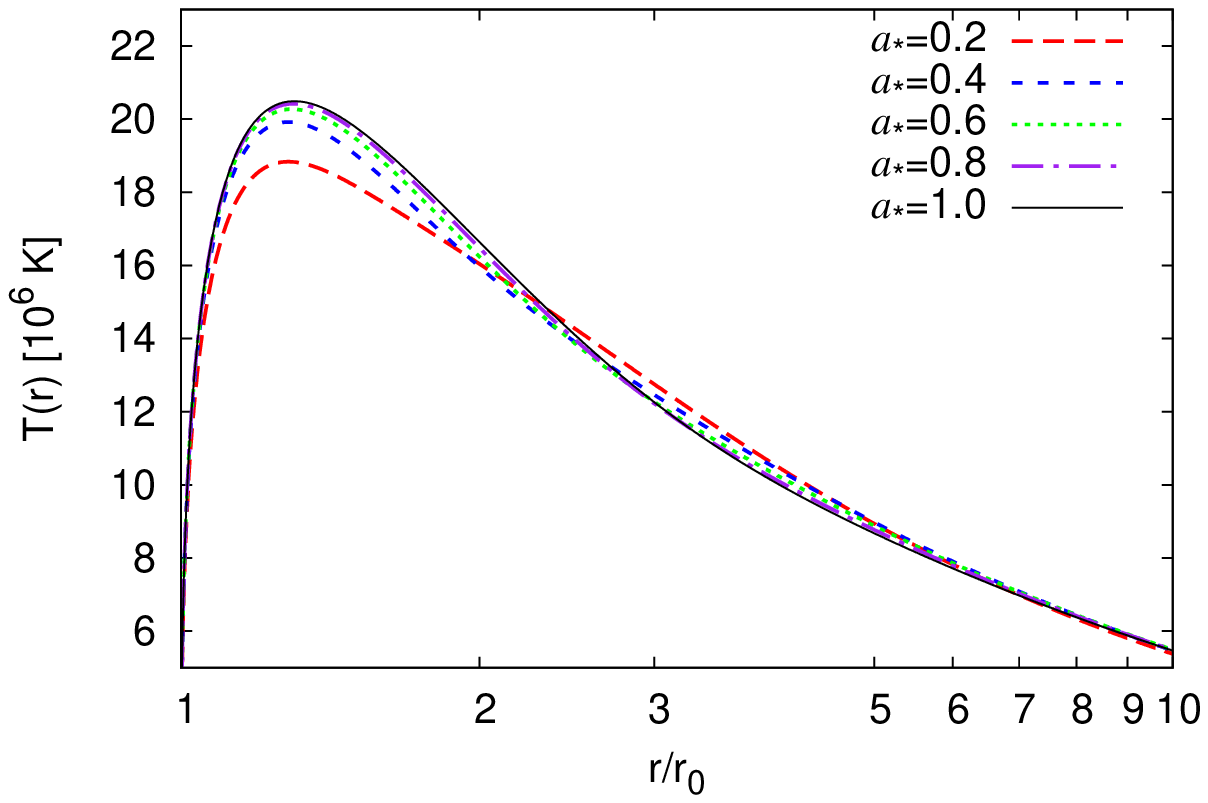}
\hspace{0.2in}
\includegraphics[width=2.8in]{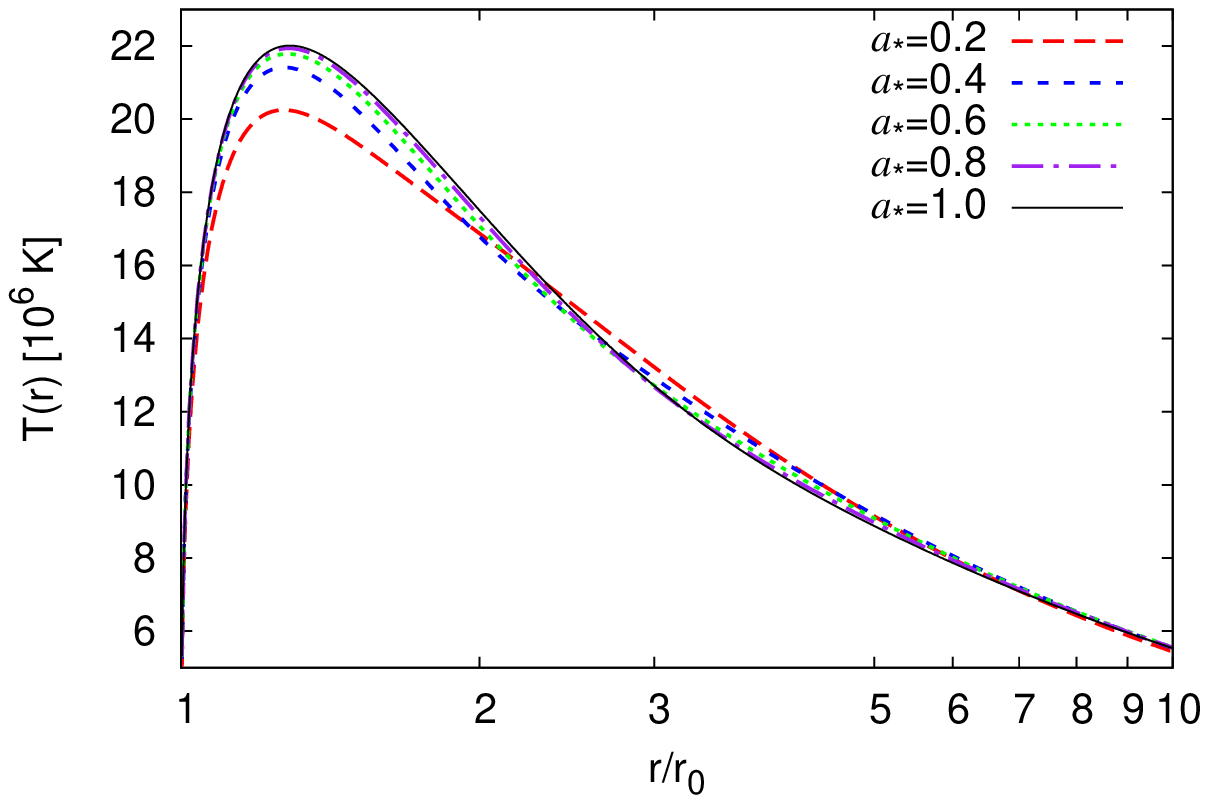}
\hspace{0.2in}
\includegraphics[width=2.8in]{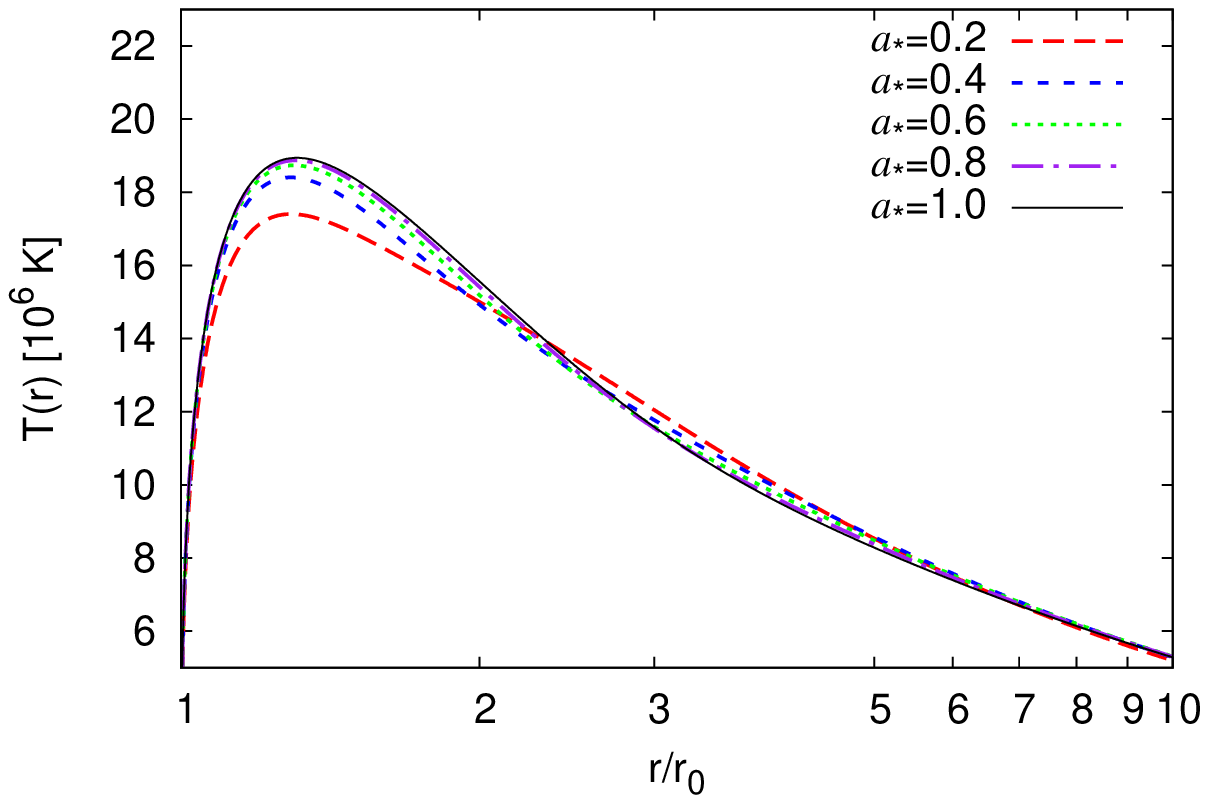}
\caption{Temperature distribution of the accretion disk in the
Kerr spacetime (upper left hand plot), and of the stationary
axially symmetric wormhole spacetimes for $\Phi=- r_0/r$, and
$b=r_0$ (upper right hand plot), $b=r^2_0/r$ (lower left hand
plot), and $b = (r_0 r)^{1/2}$ (lower right hand plot),
respectively. In all plots $r_0 =1000$ cm, and the values of the
spin parameter are $a_* = 0.2, 0.4, 0.6, 0.8$ and 1,
respectively.}
 \label{Fig:rotwh-temp}
\end{figure*}

\begin{figure*}
\centering
\includegraphics[width=2.8in]{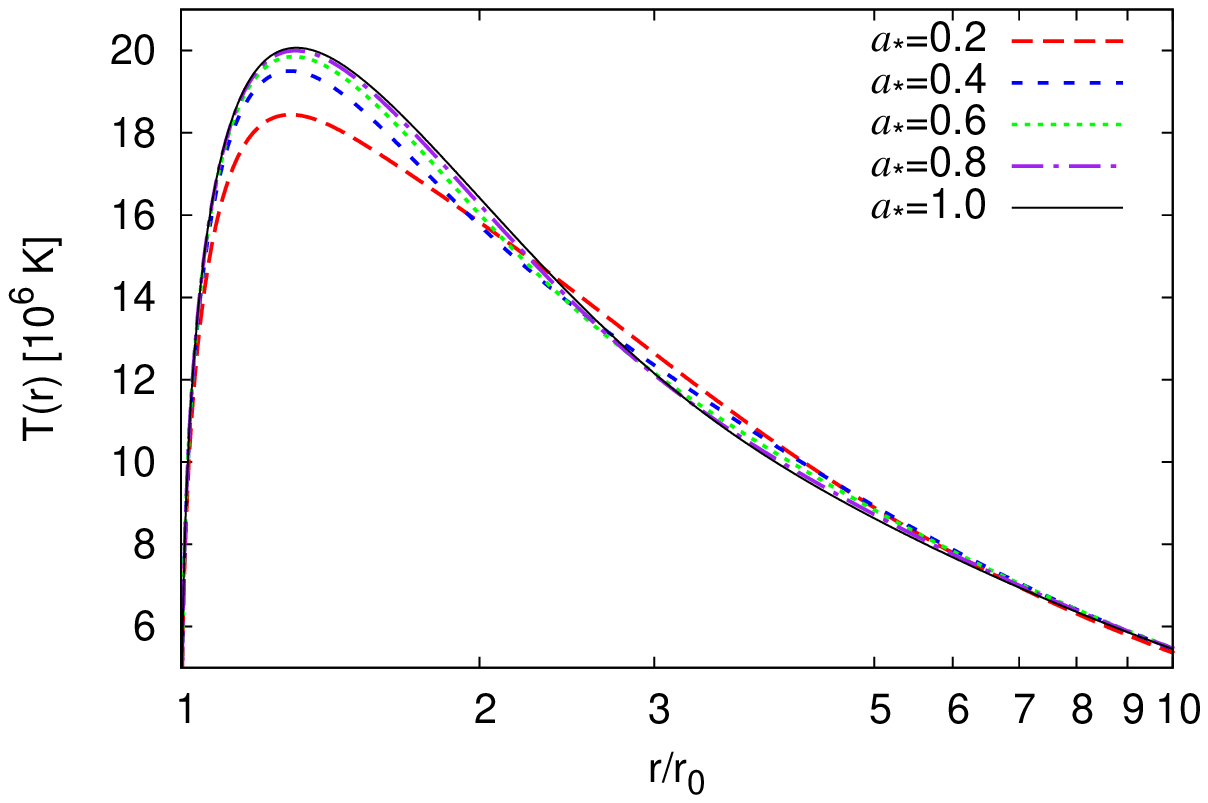}
\hspace{0.2in}
\includegraphics[width=2.8in]{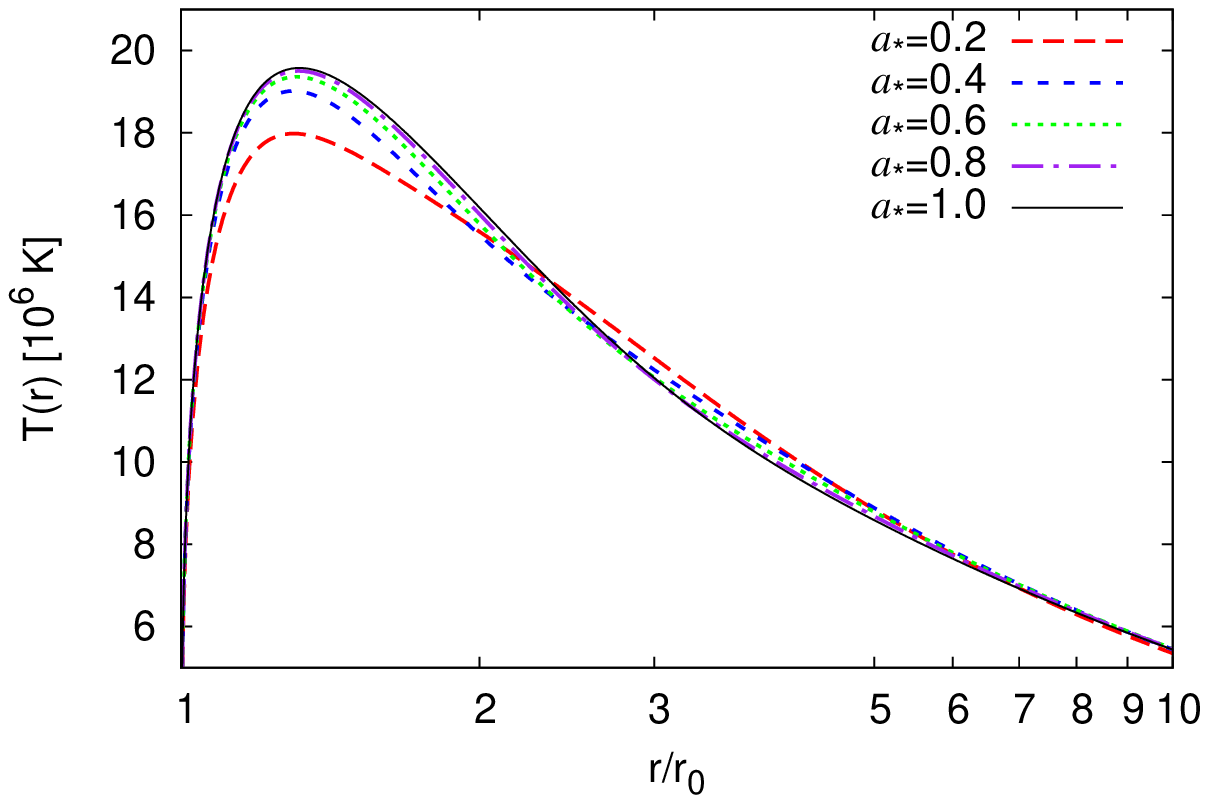}
\hspace{0.2in}
\includegraphics[width=2.8in]{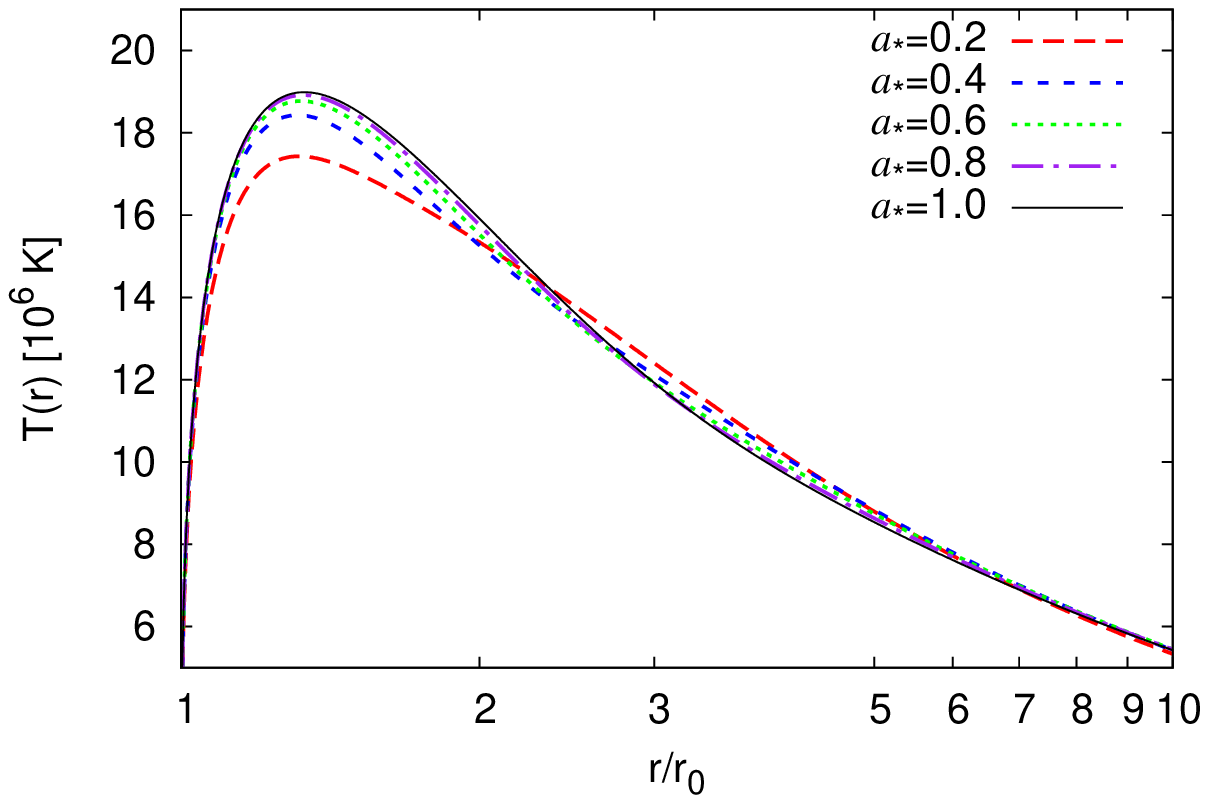}
\hspace{0.2in}
\includegraphics[width=2.8in]{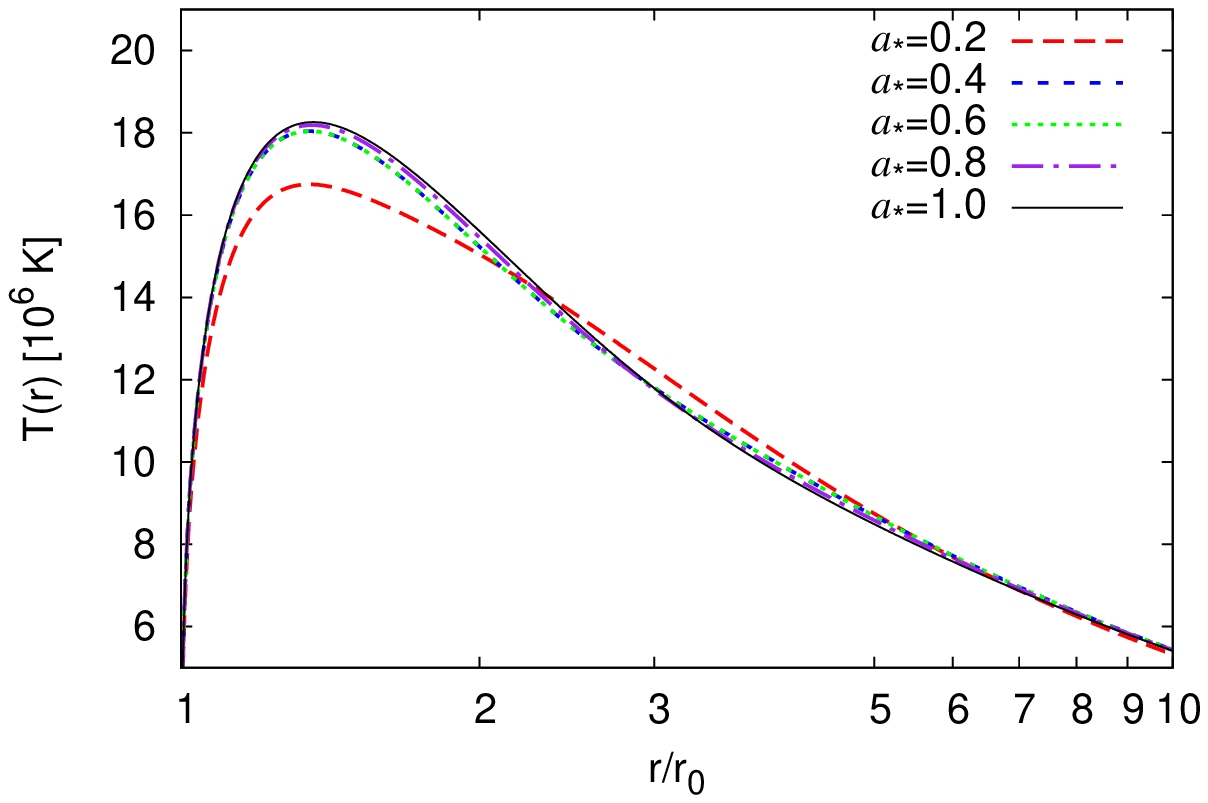}
\caption{Temperature distribution of the accretion disk in the
stationary axially symmetric wormhole spacetime for $\Phi=-
r_0/r$, and $b=r_0+\gamma r_0(1-r_0/r)$, where $\gamma=0.2$ (upper
left hand plot), $0.4$ (upper right hand plot), $0.6$ (lower left
hand plot) and $0.8$ (lower right hand plot), respectively.  In
all plots $r_0 =1000$ cm, and the values of the spin parameter are
$a_* = 0.2, 0.4, 0.6, 0.8$ and 1, respectively.}
 \label{Fig:rotwh-temp2}
\end{figure*}

\begin{figure*}
\centering
\includegraphics[width=2.8in]{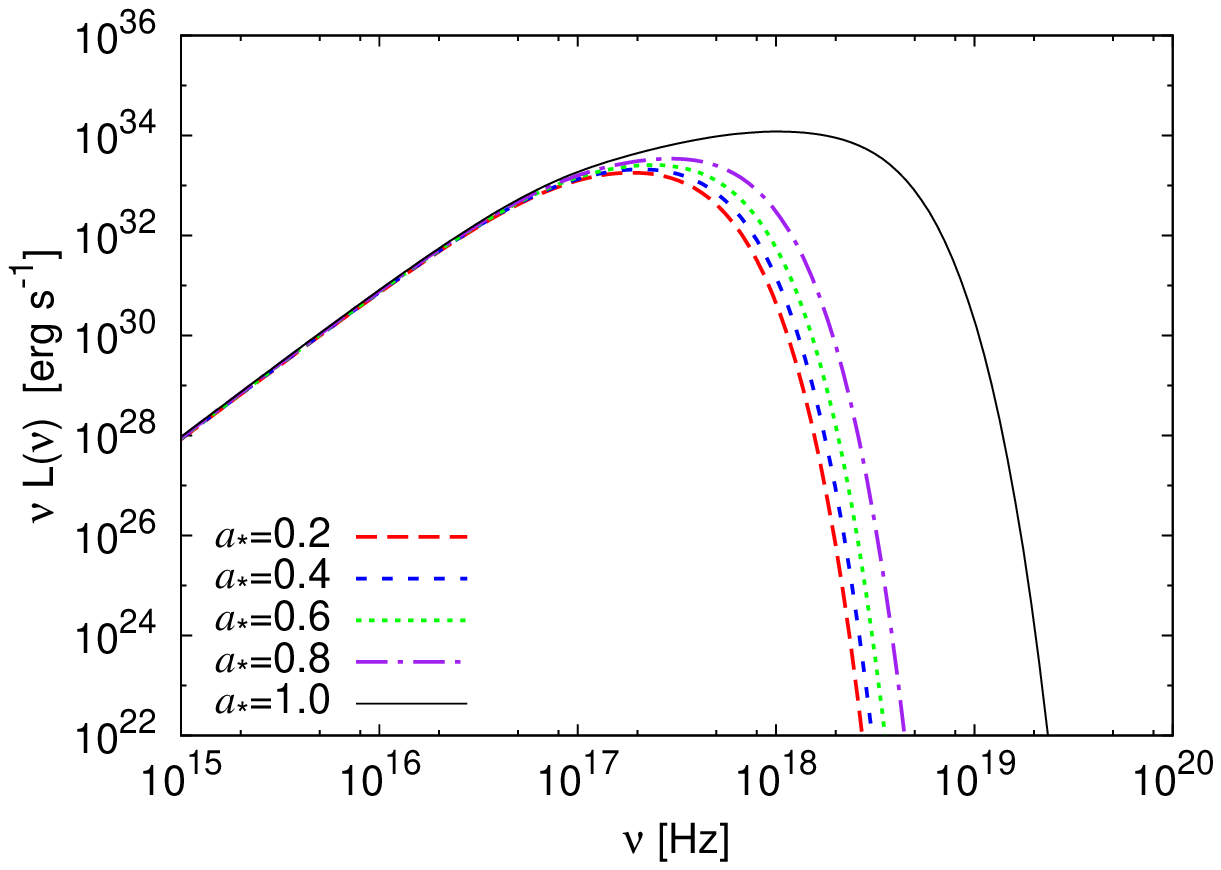}
\hspace{0.2in}
\includegraphics[width=2.8in]{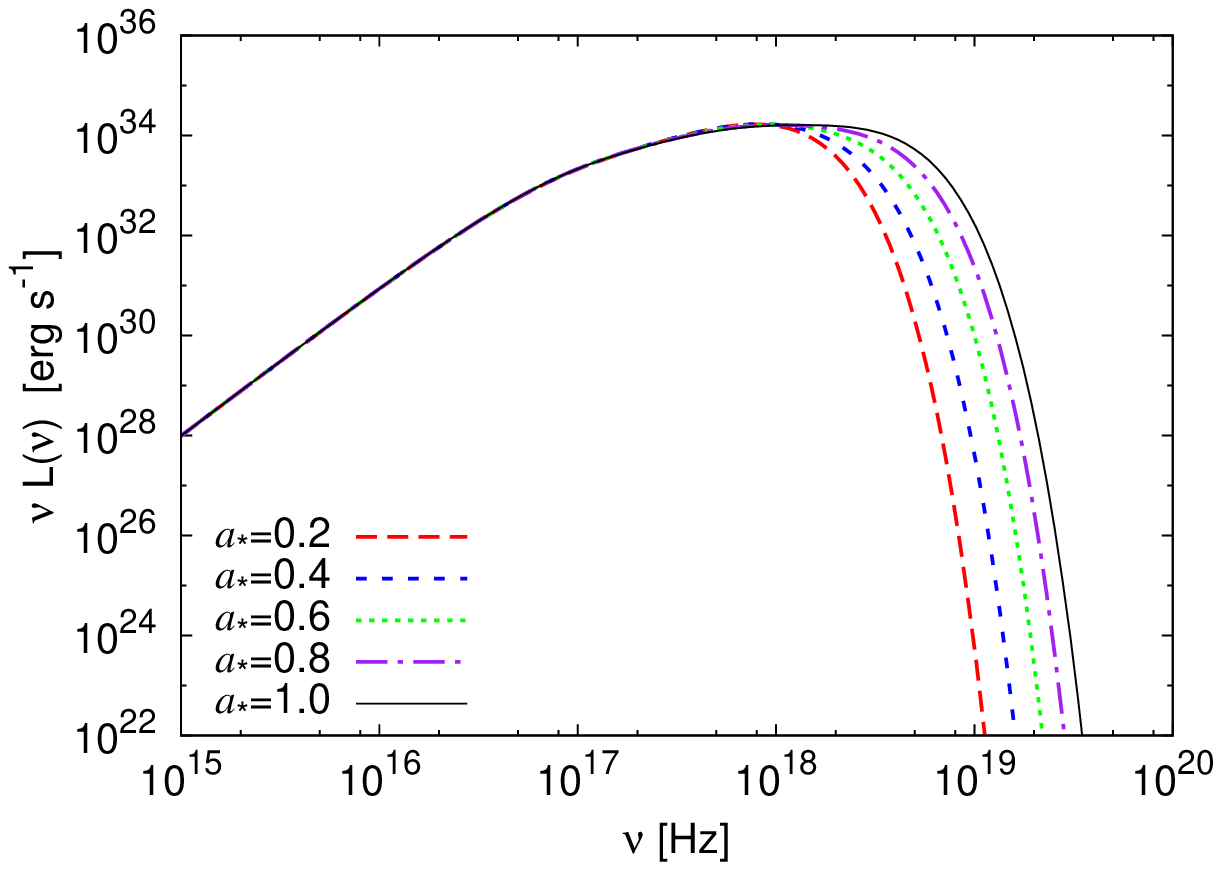}
\hspace{0.2in}
\includegraphics[width=2.8in]{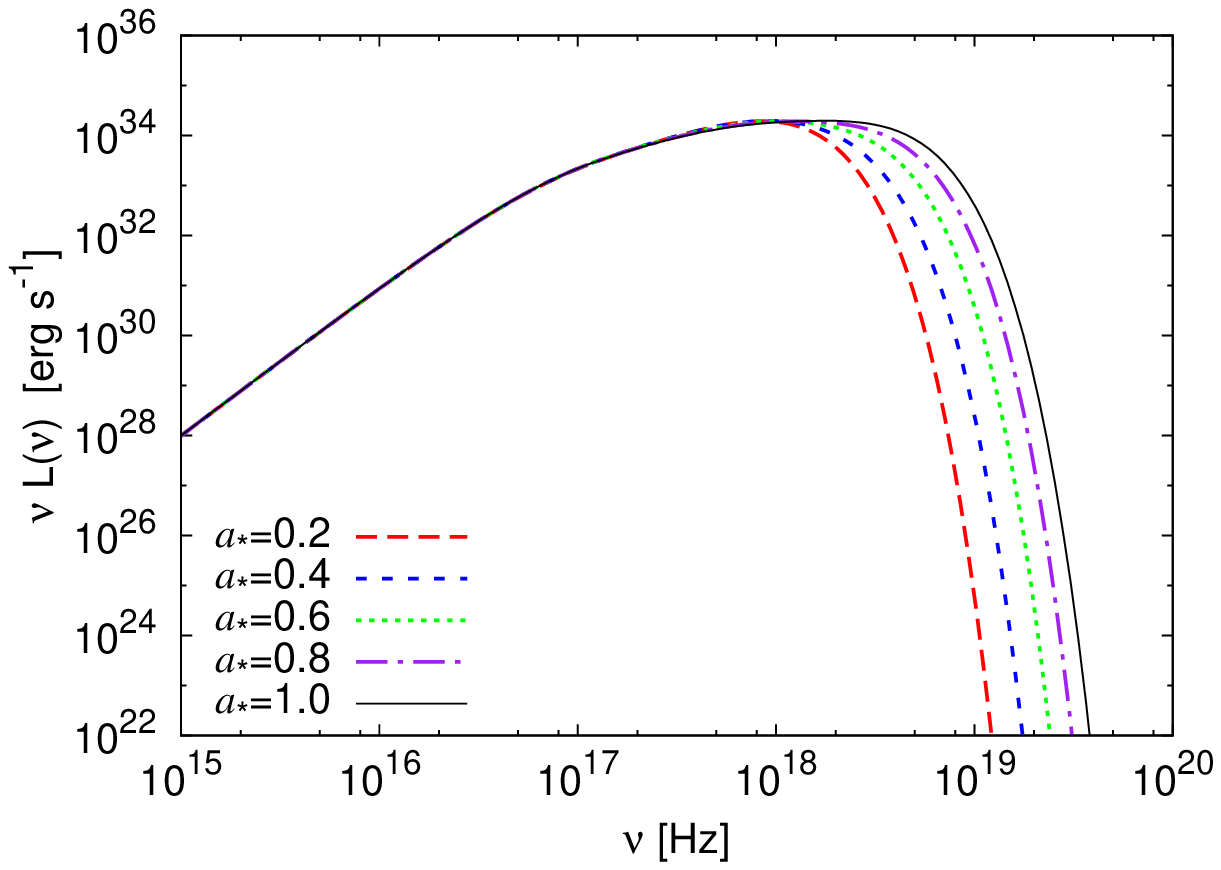}
\hspace{0.2in}
\includegraphics[width=2.8in]{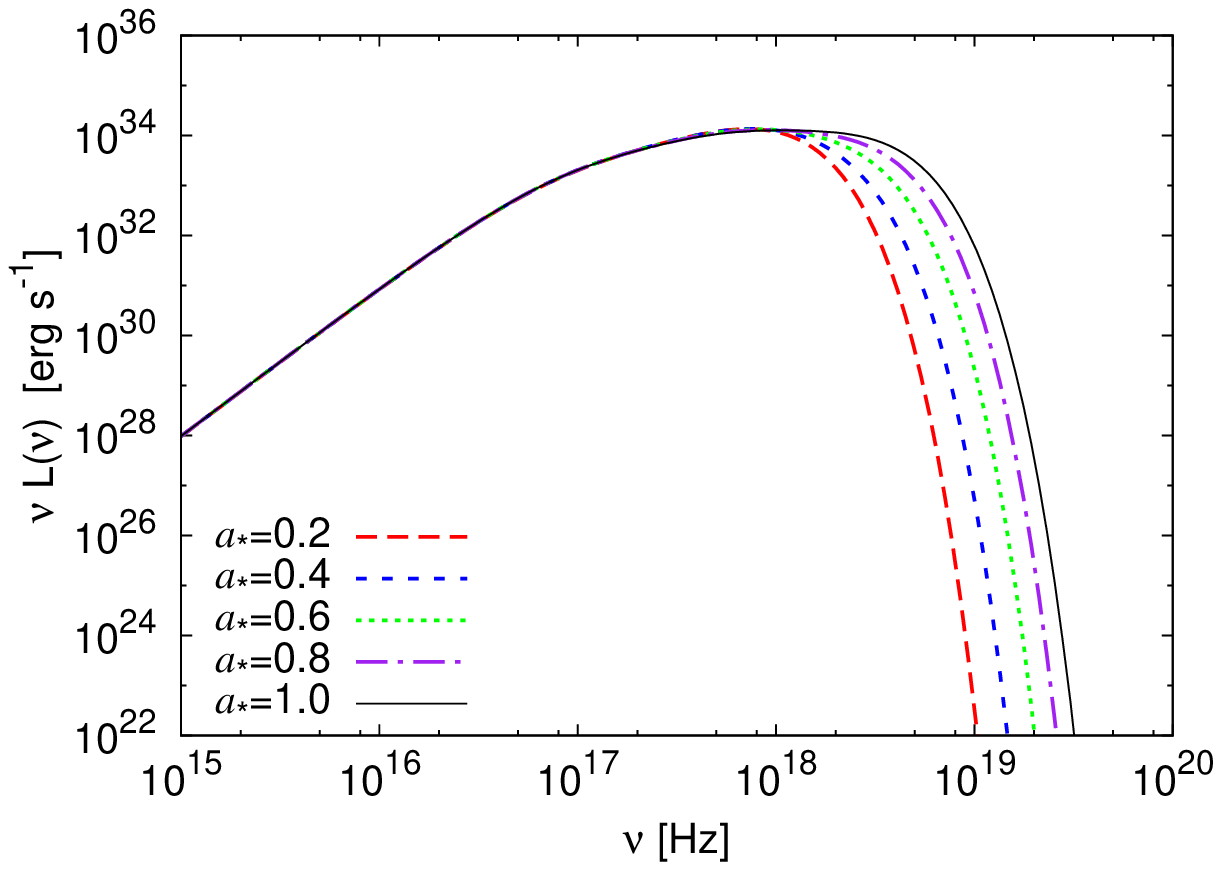}
\caption{The emission spectra of the accretion disk in the Kerr
spacetime (upper left hand plot) and in the stationary axially
symmetric wormhole spacetime for $\Phi=- r_0/r$, and $b=r_0$
(upper right hand plot), $b=r^2_0/r$ (lower left hand plot), and
$b = (r_0 r)^{1/2}$ (lower right hand plot), respectively. In all
plots $r_0 =1000$ cm, and the values of the spin parameter are
$a_* = 0.2, 0.4, 0.6, 0.8$ and 1, respectively.}
 \label{Fig:rotwh-spect}
\end{figure*}

\begin{figure*}
\centering
\includegraphics[width=2.8in]{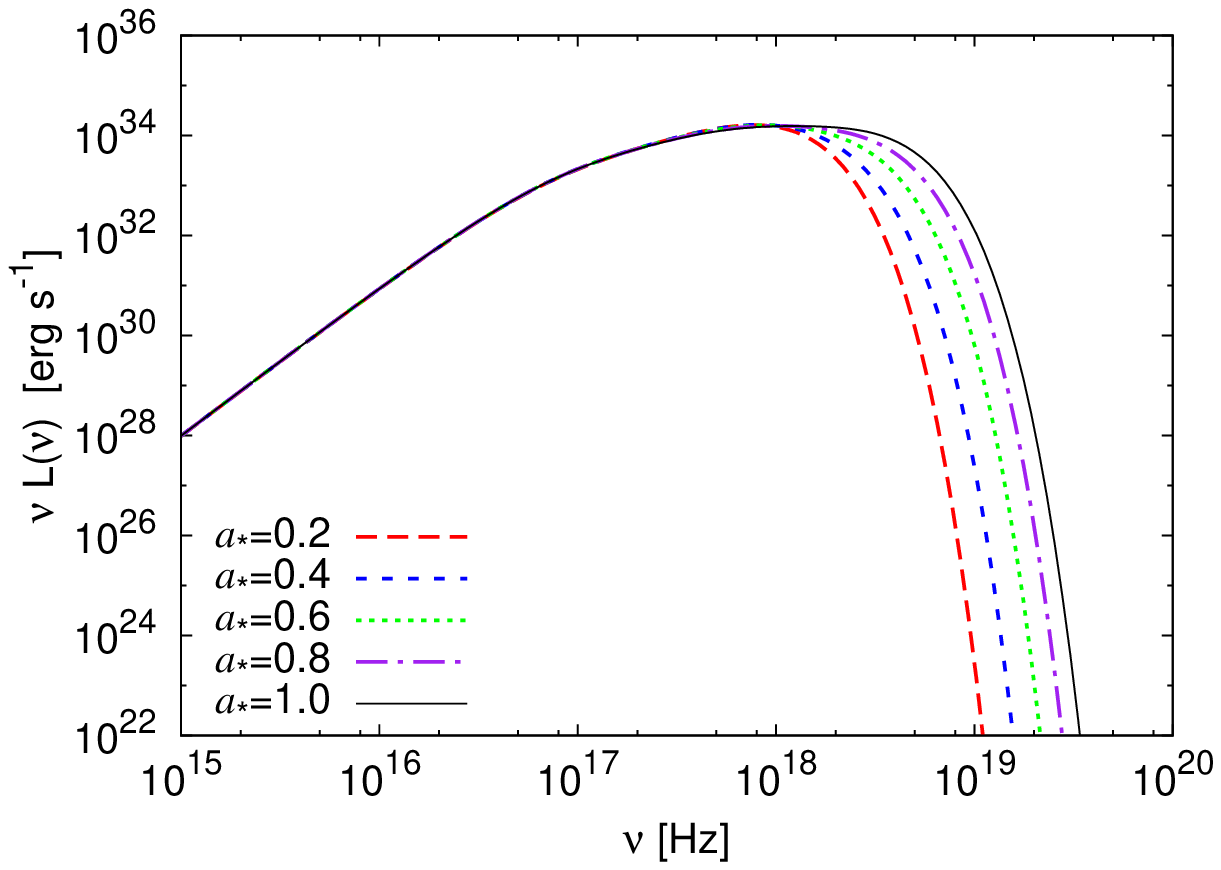}
\hspace{0.2in}
\includegraphics[width=2.8in]{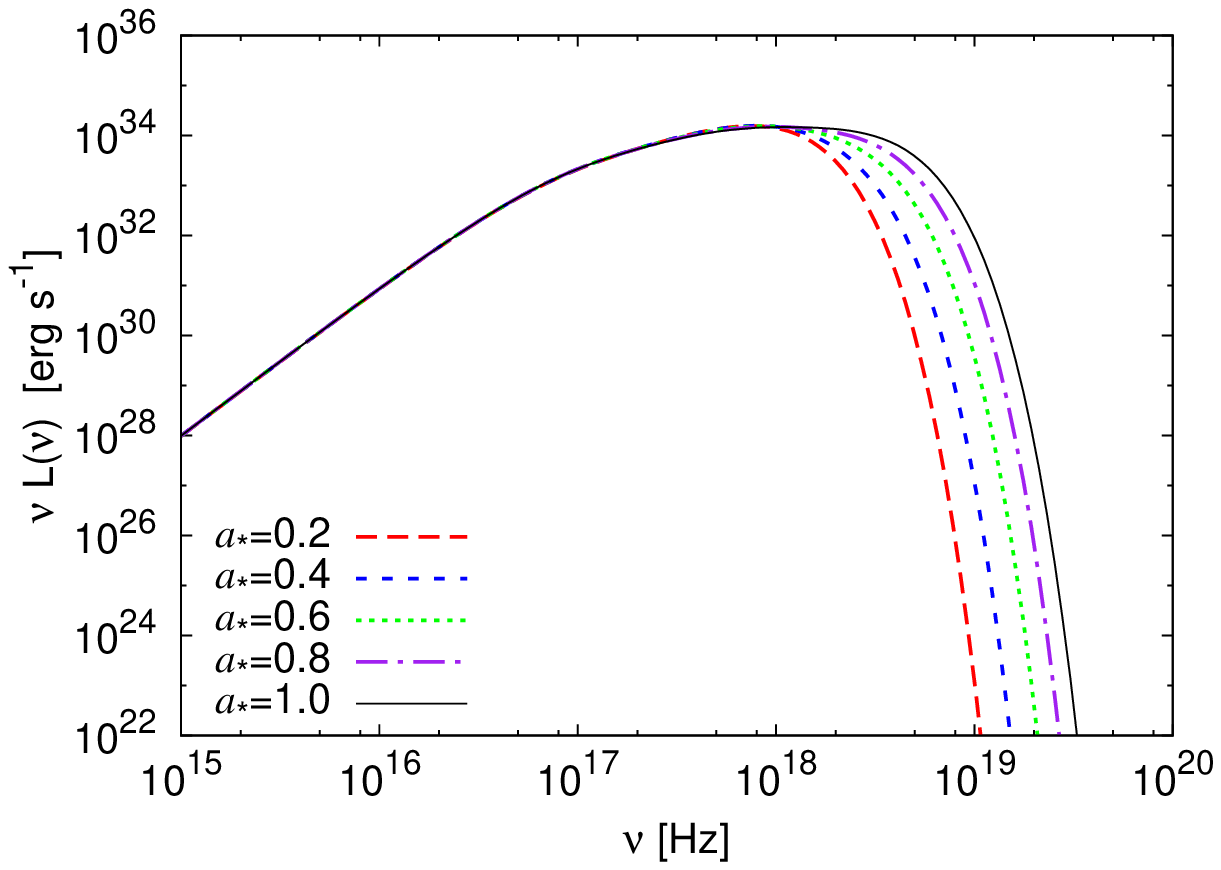}
\hspace{0.2in}
\includegraphics[width=2.8in]{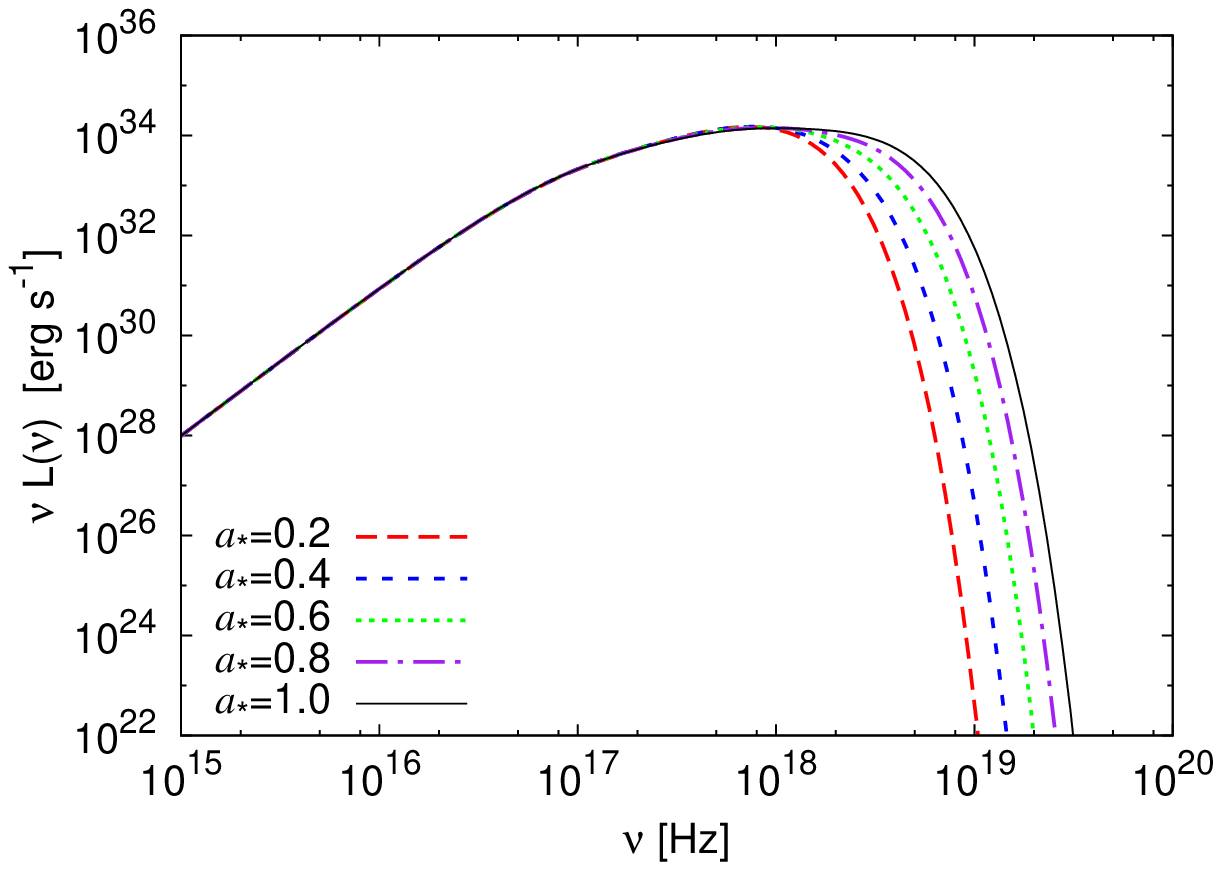}
\hspace{0.2in}
\includegraphics[width=2.8in]{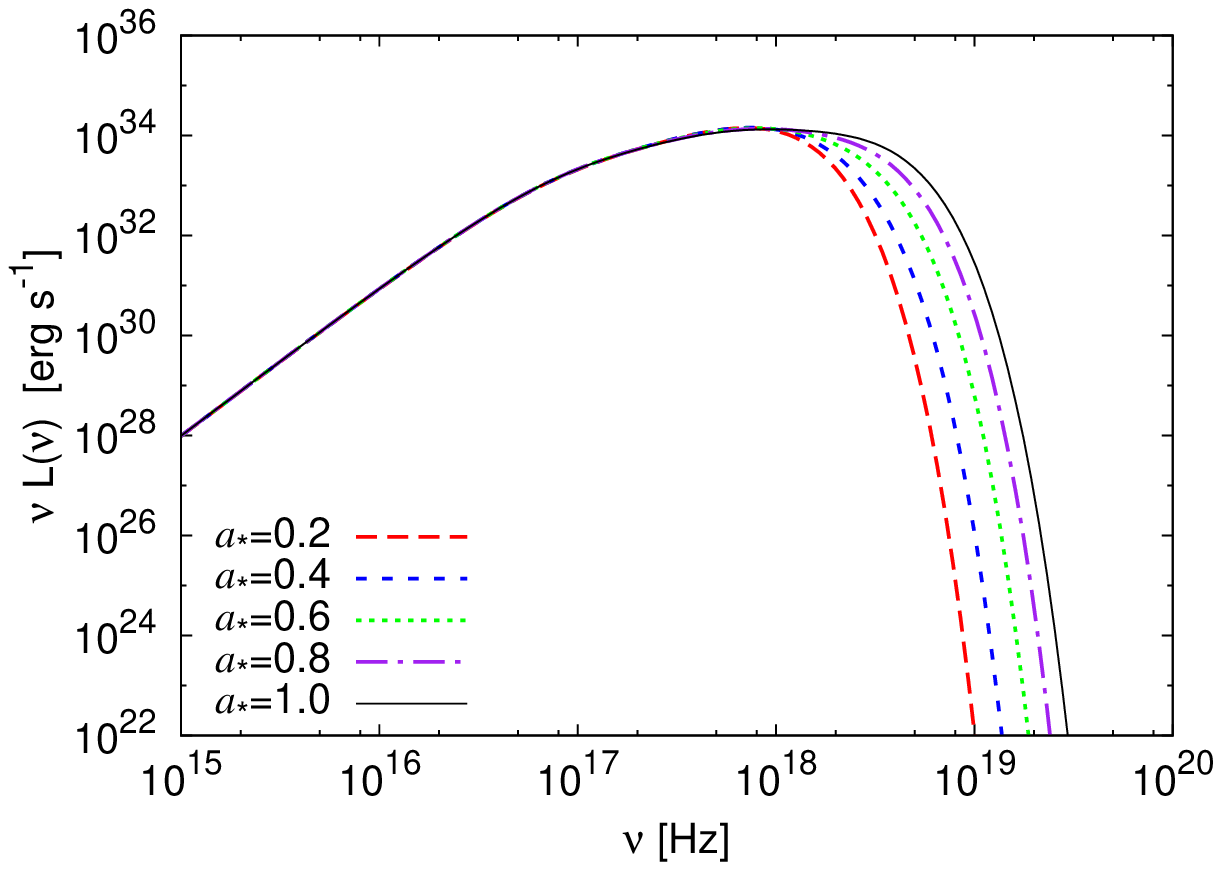}
\caption{The emission spectra of the accretion disk in the
stationary axially symmetric wormhole spacetime for $\Phi=-
r_0/r$, and $b=r_0+\gamma r_0(1-r_0/r)$, where $\gamma=0.2$ (upper
left hand plot), $0.4$ (upper right hand plot), $0.6$ (lower left
hand plot) and $0.8$ (lower right hand plot), respectively.  In
all plots $r_0 =1000$ cm, and the values of the spin parameter are
$a_* = 0.2, 0.4, 0.6, 0.8$ and 1, respectively.}
 \label{Fig:rotwh-spect2}
\end{figure*}

In Figs.~\ref{Fig:rotwh-flux}--\ref{Fig:rotwh-spect2} we plot the
energy flux, the disk temperature and the emission spectra $\nu
L(\nu)$ emitted by the accretion disk with a mass accretion rate
of $\dot{M}_0=10^{-12}M_{\odot}/{\rm yr}$ for various wormhole
geometries. The form functions we have used  are given by
Eqs.~(\ref{form1}) and (\ref{form2}), respectively. We also
present, for the sake of comparison, the properties of the thin
disks in the Kerr black hole geometry.

Comparing the energy flux, depicted in Figs. \ref{Fig:rotwh-flux}
and \ref{Fig:rotwh-flux2}, from the thin disk in a stationary
rotating wormhole geometry with the one of a Kerr black hole for
$a_*<0.8$, we see that the intensity of the flux emerging from the
disk surface is at least two orders of magnitude greater for
wormholes than for the rotating black hole with the same
geometrical mass $r_0$ and accretion rate $\dot{M}_0$. The flux
amplitude exhibits the similar dependence on the spin parameter
for both types of the rotating central objects: with the
increasing values of $a_*$, the maximal values of $\mathcal{F}(r)$
are also increasing. As the central object is rotating faster the
flux maxima are shifted closer to the black hole, whereas they are
located somewhat at lower disk radii in wormhole spacetimes. Since
each value of $a_*$ in the plots is greater than the critical spin
parameter (discussed in the next section), the inner edge of the
accretion disk, i.e., the left edge of the flux profile, is always
located at $r_0$, the throat of the wormhole, in contrast to the
different positions of the inner edge of the disk around the Kerr
black hole.

The different shape functions (\ref{form1})-(\ref{form2}) in the
wormhole metric change only the amplitudes of the flux profiles:
the accretion disk of the wormhole with the shape function
$b=r^2_0/r$ produces the radiation flux with highest intensity,
and we obtain the lowest values of $\mathcal{F}(r)$ for $b(r)=(r_0
r)^{1/2}$. By increasing the numerical value of the parameter
$\gamma $ in the shape function $b(r)=r_0+\gamma r_0(1-r_0/r)$,
the intensity of the flux is decreasing, as the $1/r$ term is
becoming more and more dominant in $b(r)$.


All these characteristics appear in the disk temperature profiles,
depicted in Figs.~\ref{Fig:rotwh-temp} and \ref{Fig:rotwh-temp2},
as well. The temperature of the disk radiation for the Kerr black
hole and the wormhole geometry have the same order of magnitude
only for $a_*\approx 1$. For lower values of the spin parameter
the disks rotating around the wormholes are much hotter than those
around the Kerr black holes.

In Figs.~\ref{Fig:rotwh-spect} and \ref{Fig:rotwh-spect2}, we
display the disk spectra for the rotating wormholes. Although
there are no significant differences between the spectral
amplitudes of the disk radiation for the different rotating
central objects, the cut-off frequencies of the spectra highly
depend on the nature of the holes and their rotational velocities.
The cut-off frequencies for the Kerr black hole are systematically
lower than those for the wormholes. However, the spectral profiles
of the radiation emitted by the accretion disk are rather similar,
no matter which of the shape functions $b(r)$ we use for the
wormhole spacetimes. The cut-off frequencies of the spectra take
the lowest values for $b(r) = (r_0 r)^{1/2}$, and the highest ones
can be found in the plot for $b(r)=r_0+\gamma r_0(1-r_0/r)$ but
these differences are negligible. The plots in
Fig.~\ref{Fig:rotwh-spect2} show that the spectral features are
also insensitive to the variation of the parameter $\gamma $ in
the shape function $b(r)=r_0+\gamma r_0(1-r_0/r)$.

%

We also present in Table~\ref{Efficiency} the conversion
efficiency $\epsilon$ of the accreting mass into radiation for the
case where the photon capture by the Kerr black hole is ignored.
The value of $\epsilon$ measures the efficiency of energy
generating mechanism by mass accretion. The amount of energy
released by matter leaving the accretion disk and falling down the
black hole or through the throat of the wormhole is the binding
energy $\widetilde{E}(r)|_{r=r_{in}}$ of the hole potential.

\begin{table}[tbp]
\begin{center}
\begin{tabular}{|c|cc|cc|}
\hline
&  $\;\;\;\;\;\;\;\;$Kerr black hole & &
$\;\;\;\;\;\;\;\;\;\;\;\;\;$ Wormhole & \\
\hline
$a_*$ & $r_{in}/r_0$ & $\epsilon$ & $r_{in}/r_0$ & $\epsilon$   \\
\hline
0.2   &   5.33       &  0.065     &     1.00     &  0.498 \\
0.4   &   4.62       &  0.075     &     1.00     &  0.506 \\
0.6   &   3.83       &  0.091     &     1.00     &  0.507 \\
0.8   &   2.91       &  0.122     &     1.00     &  0.508 \\
1.0   &   1.00       &  0.421     &     1.00     &  0.508 \\
\hline
\end{tabular}
\end{center}
\caption{The inner edge of the accretion disk and the efficiency
for rotating black hole and wormhole geometries.}
\label{Efficiency}
\end{table}

Table~\ref{Efficiency} shows that $\epsilon$ is always higher for
rotating wormholes than for Kerr black holes. Even for $a_*=0.2$
the conversion efficiency of the accretion process in the wormhole
potential is greater than $\epsilon$ derived for the extreme Kerr
black holes with $a_*=1$. If we consider rapidly rotating
wormholes, the efficiency is even higher but there is a limit of
$\epsilon$ for high values of the spin parameter: at $a_*\gtrsim
0.6$ we obtain $\epsilon=0.508$, where the accretion process is
reaching its saturation point, and we cannot attain higher
efficiencies even if we increase $a_*$ further. However, these
high numbers demonstrate that the rotating wormholes provide a
much more efficient engine for the transformation of the energy of
the accreting mass into radiation than the Kerr black holes.

\subsection{Extremely slowly rotating wormholes}\label{sec:IV}

\begin{figure*}
\centering
\includegraphics[width=2.8in]{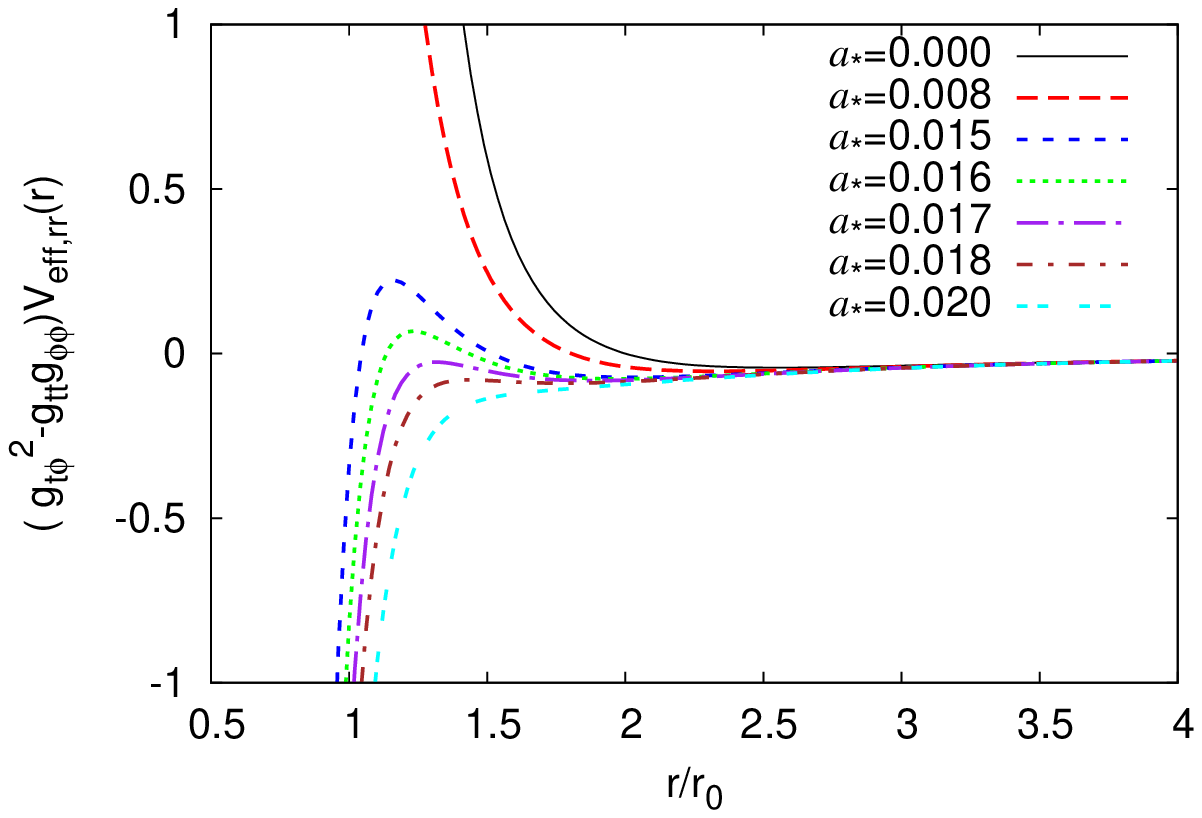}
\hspace{0.2in}
\includegraphics[width=2.8in]{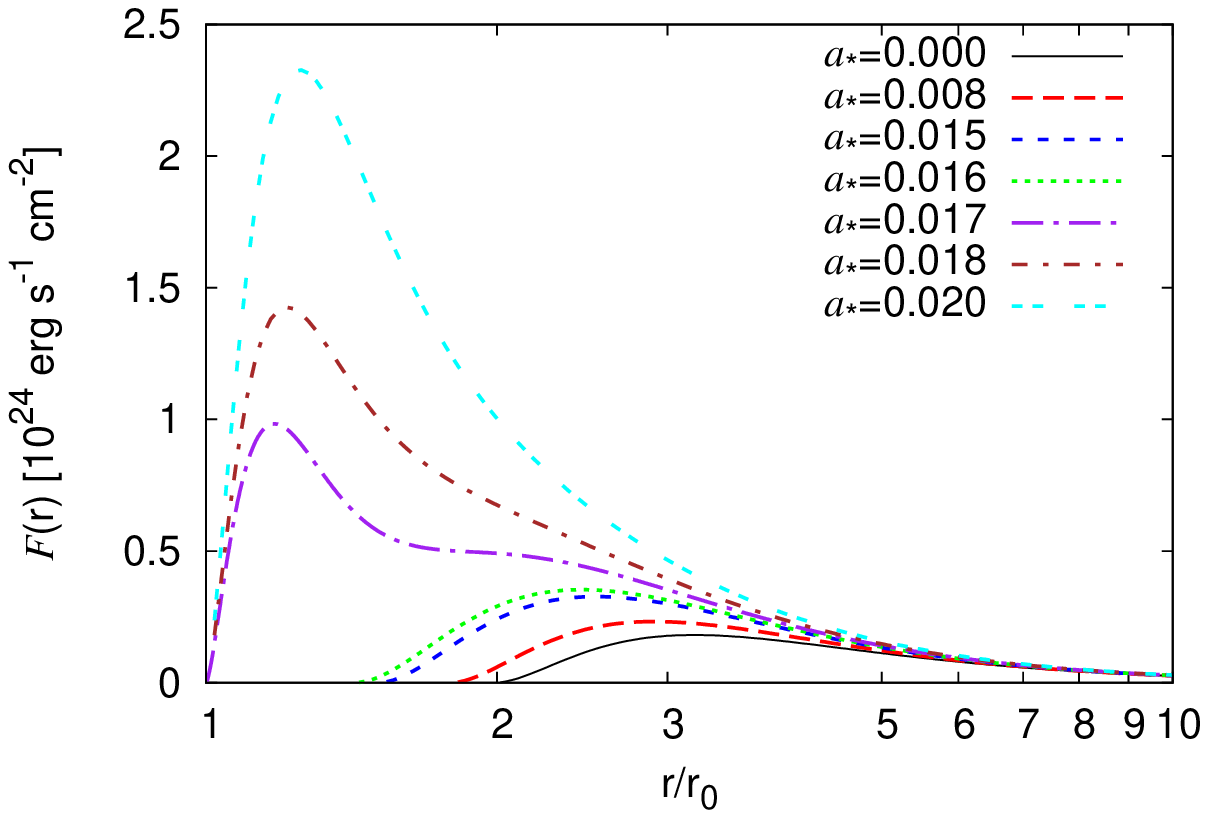}
\caption{The function $(g_{t\phi}^2-g_{tt}g_{\phi\phi})V_{eff,rr}$
vs. $r/r_0$ (left hand plot) and the flux radiated by the
accretion disk (right hand plot) in the stationary axially
symmetric wormhole spacetime for $\Phi=- r_0/r$, and $b=r_0$ for
extremely low values of the spin parameter between $0$ and
$0.02$.}
 \label{Fig:rotwh-a_star}
\end{figure*}

As seen in the previous section, the inner edge of the accretion
disk is always located at $r_0$ for any value of the spin
parameter we considered. Nevertheless, for extremely small values
of $a_*$ we found some critical behavior related to the dependence
of the location of the marginally stable orbit on the spin
parameter. By increasing $a_*$ from zero, the radius $r_{ms}$
decreases from $2r_0$. There is a critical value of the spin
parameter, namely $a_{*}^{crit}=0.016693$, where $r_{ms}=1.29r_0$
and any further small rise of $a_*$ results in a jump of $r_{ms}$
from $1.29r_0$ to $0.19 r_0$. The latter value is already behind
the throat of the wormhole, and therefore the boundary of the
allowed region for the stable circular orbits ($V_{eff,rr}$ always
remains negative) expand from $1.29 r_0$ to the throat. This
phenomenon is demonstrated in Fig. \ref{Fig:rotwh-a_star}, which
depicts the quantity given by Eq. (\ref{mso-r}) determining the
marginally stable orbit and the radial profiles of the photon flux
for small values of $a_*$.

For a value of the spin parameter between 0 and $a_{*}^{crit}$ the
function $(g_{t\phi}^2-g_{tt}g_{\phi\phi})V_{eff,rr}$ has zeros
between  $2r_{0}$ and $1.29 r_0$. For $a_*>a_{*}^{crit}$ there are
no solutions for Eq. (\ref{mso-r}), although
$(g_{t\phi}^2-g_{tt}g_{\phi\phi})V_{eff,rr}<0$ for any $r$
($g_{t\phi}^2-g_{tt}g_{\phi\phi}$ is always positive), indicating
the presence of stable circular orbits for $r>r_0$. At the
critical value of the spin parameter the flux profiles also show
the jump of the inner edge of the accretion disk from $1.29 r_0$
to $r_0$.

If we substitute the metric (\ref{3rwh}) in Eq. (\ref{mso-r}), we get
\begin{eqnarray}
0 & = & V_{eff,r} = \frac{-2e^{2\Phi}(\Phi_{,rr}+2\Phi_{,r}^{2})}
{-e^{2\Phi}+r^{2}K^{2}(\omega-\Omega)^{2}}\nonumber\\
& & +\frac{2K^{2}[(\omega+r\omega_{,r})^{2}+r\omega(2\omega_{,r}
+r\omega_{,rr})]}{-e^{2\Phi}+r^{2}K^{2}(\omega-\Omega)^{2}}
   \nonumber\\
& &  +\frac{-2K^{2}\Omega[2\omega+r(4\omega_{,r}+r\omega_{,rr})]
+2K^{2}\Omega^{2}}{-e^{2\Phi}+r^{2}K^{2}(\omega-\Omega)^{2}}
     \nonumber\\
& & +2(Kr\omega_{,r}e^{-\Phi})^{2}+8r^{-1}\Phi_{,r}\;,\label{V_eff}
\end{eqnarray}
which is a general formula for any stationary axially symmetric
wormhole. Here
\begin{eqnarray*}
\Omega & = &
\frac{1}{2}(2\omega+r\omega_{,r})+\frac{1}{2}\sqrt{r^{2}
\omega_{,r}^{2}+4K^{-2}r^{-1}\Phi_{,r}e^{2\Phi}}\;.
\end{eqnarray*}
For the specific case in study, expression (\ref{V_eff}) leads to
the following relationship
\begin{eqnarray*}
0 & = &
 \frac{e^{-2/x^{2}}\left[x^{4}+2(x^{2}-1)\right]
 +(6a_{*}x^{-2})^{2}}{-x^{4
}e^{-2/x^{2}}+4a_{*}-\sqrt{9a_{*}^{2}+x^{6}e^{-2/x^{2}}}}
   \\
& & +1+(6a_{*}x^{-4}e^{1/x^ {2}})^{2}-4x^{-2}\;,
\end{eqnarray*}
where $x^2=r/r_0$. This result shows that the location of the
marginally stable orbit given in the dimensionless radial
coordinate $x^2$ depends only on $a_*$ and the value of
$a_{*}^{crit}$ does not depend on $r_0$, as depicted in Fig.
\ref{Fig:rotwh-a_star}.

By comparing two states with the spin parameters negligibly
smaller and greater than $a_*^{crit}$ we see that the distance
between the inner edge of the accretion disk and the wormhole
throat is $1.29r_0-r_0=0.29 r_0$ in both cases. If the accretion
process increases the spin of the wormhole such that it just
exceeds $a_{*}^{crit}$ with a infinitesimally small value then the
inner edge of the disk will fall into the throat with supersonic
velocity, eliminating the so called plunging region located
between the disk and the rotating central object. It must have a
strong physical influence, e.g., shock waves are generated.
However, after this transient situation, the inner edge of the
disk will be located at $r_0$ and all the orbits for $r>r_0$ will
be stable $(V_{eff,rr}<0)$, i.e., the thin accretion disk exists
in the whole equatorial plane outside the throat.

\section{Discussions and final remarks}\label{sec:concl}

In the present paper, we have studied thin accretion disk models
in stationary axially symmetric wormhole geometries, and have
carried out an analysis of the properties of the radiation
emerging from the surface of the disk. In classical general
relativity, wormholes are supported by exotic matter, which
involves a stress-energy tensor that violates the null energy
condition. Thus, it is important to analyze the properties of the
accretion disks around wormholes supported by exotic matter,
namely, the time averaged energy flux, the disk temperature and
the emission spectra. By comparing the accretion disk properties
in a stationary rotating wormhole geometry with the properties of
disks around a Kerr black hole, we have shown that the intensity
of the flux emerging from the disk surface is greater for
wormholes than for the rotating black hole with the same
geometrical mass $r_0$ and accretion rate $\dot{M}_0$. We also
presented the conversion efficiency $\epsilon$ of the accreting
mass into radiation, and proved that rotating wormholes are much
more efficient in converting the accreting mass into radiation
than the Kerr black holes.

It is generally expected that most of the astrophysical objects
grow substantially in mass via accretion. Recent observations
suggest that around most of the active galactic nuclei (AGN's) or
black hole candidates there exist gas clouds surrounding the
central far object, and an associated accretion disk, on a variety
of scales from a tenth of a parsec to a few hundred parsecs
\cite{UrPa95}. These clouds are assumed to form a geometrically
and optically thick torus (or warped disk), which absorbs most of
the ultraviolet radiation and the soft x-rays. The gas exists in
either the molecular or the atomic phase. The most powerful
evidence for the existence of super massive black holes comes from
the very long baseline interferometry (VLBI) imaging of molecular
${\rm H_2O}$ masers in the active galaxy NGC 4258 \cite{Miyo95}.
This imaging, produced by Doppler shift measurements assuming
Keplerian motion of the masering source, has allowed a quite
accurate estimation of the central mass, which has been found to
be a $3.6\times 10^7M_{\odot }$ super massive dark object, within
$0.13$ parsecs. Hence, important astrophysical information can be
obtained from the observation of the motion of the gas streams in
the gravitational field of compact objects.

The determination of the accretion rate for an astrophysical
object can give a strong evidence for the existence of a surface
of the object. A model in which Sgr A*, the $3.7\times 10^6
M_{\odot }$ super massive black hole candidate at the Galactic
center, may be a compact object with a thermally emitting surface
was considered in \cite{BrNa06}. For very compact surfaces within
the photon orbit, the thermal assumption is likely to be a good
approximation because of the large number of rays that are
strongly gravitationally lensed back onto the surface. Given the
very low quiescent luminosity of Sgr A* in the near-infrared, the
existence of a hard surface, even in the limit in which the radius
approaches the horizon, places a severe constraint on the steady
mass accretion rate onto the source, ${\dot M}\le 10^{-12}
M_{\odot}$ yr$^{-1}$. This limit is well below the minimum
accretion rate needed to power the observed submillimeter
luminosity of Sgr A*, ${\dot M}\ge 10^{-10} M_{\odot}$ yr$^{}$.
Thus, from the determination of the accretion rate it follows that
Sgr A* does not have a surface, that is, it must have an event
horizon. Therefore the study of the accretion processes by compact
objects is a powerful indicator of their physical nature. However,
up to now, the observational results have confirmed the
predictions of general relativity mainly in a qualitative way.
With the present observational precision one cannot distinguish
between the different classes of compact/exotic objects that
appear in the theoretical framework of general relativity
\cite{YuNaRe04}.

However, important technological developments may allow one to
image black holes and other compact objects directly
\cite{BrNa06}. For a black hole embedded in an accretion flow, the
silhouette will generally be asymmetric regardless of the spin of
the black hole. Even in an optically thin accretion flow an
asymmetry will result from special relativistic effects
(aberration and Doppler shifting). In principle, detailed
measurements of the size and shape of the silhouette could yield
information about the mass and spin of the central object, and
provide invaluable information on the nature of the accretion
flows in low luminosity galactic nuclei. With the improvement of
the imaging observational techniques,  which give the
physical/geometrical properties of the silhouette of the compact
object cast upon the accretion flows, it will also be possible to
provide clear observational evidence for the existence of
wormholes, and to differentiate them from other types of compact
general relativistic objects.

In this context, we conclude our study by pointing out that the
specific properties that appear in the physical characteristics of
the thin accretion disks around wormholes can lead to the
possibility of directly detecting and discriminating wormhole
geometries by observing accretion disks around compact
astrophysical objects.

\section*{Acknowledgments}

The work of TH is supported by an RGC grant of the government of
the Hong Kong SAR. ZK was supported by the Hungarian Scientific
Research Fund (OTKA) grant no.69036. FSNL was partially funded by
Funda\c{c}\~{a}o para a Ci\^{e}ncia e a Tecnologia (FCT)--Portugal
through the grant SFRH/BPD/26269/2006.

\appendix


\section{Effective potential for the Kerr black hole}\label{app:I}

The Kerr metric in the Boyer-Lindquist (BL) coordinate system is given by
\begin{eqnarray}
ds^{2}&=& -\left(1-\frac{2mr}{\Sigma}\right)dt^{2}+2\frac{2mr}
{\Sigma}a\sin^{2}\theta dt d\phi+\frac{\Sigma}{\Delta}dr^{2}
   \nonumber \\
&&\hspace{-0.5cm}+\Sigma d\theta^{2}+\left(r^{2}+a^{2}
+\frac{2mr}{\Sigma}a^{2}\sin^{2}\theta\right)\sin^{2}\theta
d\phi^{2}.
\end{eqnarray}
In the equatorial plane, the metric components reduce to
\begin{eqnarray*}
g_{tt} & = & -\left(1-\frac{2mr}{\Sigma}\right)=
-\left(1-\frac{2m}{r}\right),\\
g_{t\phi} & = & \frac{2mr}{\Sigma}a\sin^{2}\theta=2\frac{ma}{r}\;,\\
g_{rr} & = & \frac{\Sigma}{\Delta}=\frac{r^{2}}{\Delta},\\
g_{\phi\phi} & = & \left(r^{2}+a^{2}
+\frac{2mr}{\Sigma}a^{2}\sin^{2}\theta\right)\sin^{2}\theta\\
 & = & r^{2}+a^{2}\left(1+\frac{2m}{r}\right),
\end{eqnarray*}
respectively.

The geodesic equation (\ref{geodeqs3}) for $r$ is
\begin{equation}
\frac{r^{2}}{\Delta}\left(\frac{dr}{d\tau}\right)^{2}=V_{eff}(r),
\end{equation}
with the effective potential given by
\begin{eqnarray}
V_{eff}(r)= -1+\Big\{\widetilde{E}^{2}\left[r^{2}(r^{2}+a^{2})
 +2ma^{2}r\right]
   \nonumber  \\
 +4\widetilde{E}\widetilde{L}mar-\widetilde{L}^{2}
 \left(r^{2}-2mr\right)\Big\}\big/
 \left[r^{2}(g_{t\phi}^{2}-g_{tt}g_{\phi\phi})\right].
\end{eqnarray}
Note that these relationships may be rewritten as
\begin{equation}
r^4\left(\frac{dr}{d\tau}\right)^{2}=V(r),
  \label{KerrPot}
\end{equation}
with $V(r)$ given by
\begin{equation}
V(r)=r^{2}\Delta
V_{eff}(r)=r^{2}(g_{t\phi}^{2}-g_{tt}g_{\phi\phi})
V_{eff}(r),\label{Veff}
\end{equation}
where the relationship
$\Delta=g_{t\phi}^{2}-g_{tt}g_{\phi\phi}=r^2-2mr+a^2$ along the
equatorial plane has been used.



\begin{thebibliography}{99}

\bibitem{Harko:2008vy}
  T.~Harko, Z.~Kovacs and F.~S.~N.~Lobo,
  Phys.\ Rev.\  D {\bf 78}, 084005 (2008)
  [arXiv:0808.3306 [gr-qc]].

\bibitem{Bom}
S.~Bhattacharyya, A.~V.~Thampan and  I.~Bombaci, Astron.
Astrophys. {\bf 372}, 925 (2001).

\bibitem{To02}
D.~Torres, Nucl. Phys. B {\bf 626}, 377 (2002).

\bibitem{YuNaRe04}
Y.~ F.~Yuan, R.~Narayan and M.~J.~Rees,  Astrophys. J. {\bf 606},
1112 (2004).

\bibitem{Guzman:2005bs}
  F.~S.~Guzman,
  Phys.\ Rev.\  D {\bf 73}, 021501 (2006).

\bibitem{Pun:2008ae}
  C.~S.~J.~Pun, Z.~Kovacs and T.~Harko, Phys. Rev. D {\bf 78}, 024043
  (2008).

\bibitem{Pun:2008ua}
  C.~S.~J.~Pun, Z.~Kovacs and T.~Harko,
  Phys. Rev. D {\bf 78}, 084015 (2008).



\bibitem{NoTh73}
I.~D.~Novikov and K.~S.~Thorne, in Black
Holes, ed. C. DeWitt and B. DeWitt, New York: Gordon and Breach
(1973).

\bibitem{ShSu73}
N.~I.~Shakura and R.~A.~Sunyaev, Astron. Astrophys. {\bf%
24}, 33 (1973).

\bibitem{PaTh74}
D.~N.~Page and K.~S.~Thorne, Astrophys. J. {\bf 191}, 499 (1974).

\bibitem{Th74}
K.~S.~Thorne, Astrophys. J. {\bf 191}, 507 (1974).

\bibitem{WHsolutions}
  M.~S.~Morris and K.~S.~Thorne,
  Am.\ J.\ Phys.\  {\bf 56}, 395 (1988);
  %
H.~G.~Ellis,
  J.\ Math.\ Phys.\  {\bf 14} (1973) 104;
K.~A.~Bronnikov,
  Acta Phys.\ Polon.\  B {\bf 4} (1973) 251;
%
G.~Clement,
  Gen.\ Rel.\ Grav.\  {\bf 16}, 131 (1984);
G.~Clement,
  Gen.\ Rel.\ Grav.\  {\bf 16}, 477 (1984);
G.~Clement,
  Gen.\ Rel.\ Grav.\  {\bf 16}, 491 (1984);
  %
B.~Bhawal and S.~Kar, Phys. Rev. D {\bf 46}, 2464-2468 (1992);
%
G.~Dotti, J.~Oliva, and R.~Troncoso, Phys. Rev. D {\bf 75}, 024002
(2007);
%
L.~A.~Anchordoqui and S.~E.~P.~Bergliaffa, Phys. Rev. D {\bf 62},
067502 (2000);
%
K.~A.~Bronnikov and S.-W. Kim, Phys. Rev. D {\bf 67}, 064027
(2003);
%
M.~La~Camera, Phys. Lett. {\bf B573}, 27-32 (2003);
%
F.~S.~N.~Lobo,
  Phys.\ Rev.\ D {\bf 75}, 064027 (2007);
  %
K.~K.~Nandi, B.~Bhattacharjee, S.~M.~K.~Alam and J.~Evans,
  Phys.\ Rev.\  D {\bf 57}, 823 (1998);
  %
R.~Garattini and F.~S.~N.~Lobo,
  Class.\ Quant.\ Grav.\  {\bf 24}, 2401 (2007);
  R.~Garattini and F.~S.~N.~Lobo,
  Phys.\ Lett.\  B {\bf 671}, 146 (2009);
  %
  C.~G.~Boehmer, T.~Harko and F.~S.~N.~Lobo,
  Phys.\ Rev.\  D {\bf 76}, 084014 (2007);
%
C.~G.~Boehmer, T.~Harko and F.~S.~N.~Lobo,
  Class.\ Quant.\ Grav.\  {\bf 25}, 075016 (2008);
  A.~DeBenedictis, R.~Garattini and F.~S.~N.~Lobo,
  Phys.\ Rev.\  D {\bf 78}, 104003 (2008);
  %
S.~Sushkov, Phys. Rev. D {\bf 71}, 043520 (2005);
%
F.~S.~N.~Lobo, Phys.\ Rev.\ D {\bf 71}, 084011 (2005);
%
  F.~S.~N.~Lobo, Phys.\ Rev.\ D {\bf 71}, 124022 (2005);
%
F.~S.~N.~Lobo, Phys.\ Rev.\ D {\bf 73}, 064028 (2006);
  %
  F.~S.~N.~Lobo, Phys.\ Rev.\  D {\bf 75}, 024023 (2007);
  %
  F.~S.~N.~Lobo, Class.\ Quant.\ Grav.\  {\bf 25}, 175006 (2008);
  %
  J.~P.~S.~Lemos, F.~S.~N.~Lobo and S.~Quinet de Oliveira,
  Phys.\ Rev.\  D {\bf 68}, 064004 (2003);
%
J.~P.~S.~Lemos and F.~S.~N.~Lobo,
  Phys.\ Rev.\  D {\bf 69}, 104007 (2004);
%
H.~Maeda and M.~Nozawa,
  Phys.\ Rev.\  D {\bf 78}, 024005 (2008);
%
H.~Maeda, to appear in Physical Review D,
  arXiv:0811.2962 [gr-qc];
%
H.~Maeda, T.~Harada and B.~J.~Carr,
  arXiv:0901.1153 [gr-qc].


\bibitem{Lobo:2007zb}
  F.~S.~N.~Lobo,
  arXiv:0710.4474 [gr-qc].
\bibitem{teo}
  E.~Teo,
  Phys.\ Rev.\  D {\bf 58}, 024014 (1998).

\bibitem{Lu79}
J.~P.~Luminet, Astron. Astrophys. {\bf 75}, 228 (1979).

\bibitem{BMT01}
S.~Bhattacharyya, R.~Misra, and A.~V.~Thampan, Astrophys. J. {\bf 550}, 841  (2001).

\bibitem{UrPa95} C. M. Urry  and P. Padovani, Publ. Astron. Soc. of the Pacific {\bf 107},
803 (1995).

\bibitem{Miyo95} M. Miyoshi, J. Moran, J. Herrnstein, L. Greenhill, N. Nakai, P. Diamond  and M. Inoue,
Nature {\bf 373}, 127 (1995).

\bibitem{BrNa06}
A. E. Broderick and R. Narayan, Astrophys. JL. {\bf 636}, L109 (2006).

\end{thebibliography}
\end{document}